\documentclass{aa}
\usepackage{graphicx}
\usepackage{amssymb}
\usepackage{amsmath}
\usepackage{txfonts}
\usepackage{gensymb}
\usepackage{lscape}
\usepackage{multicol}
\usepackage{subfigure}
\usepackage{soul}
\bibpunct{(}{)}{;}{a}{}{,}

\def\phn{\phantom{0}}

\authorrunning{Magee et al.}
\titlerunning{He in SNe Iax}

\begin{document}
            
    \title{Detecting the signatures of helium in type Iax supernovae} 
	\author{M. R. Magee \inst{1}
            \and
            S. A. Sim \inst{1}
            \and
            R. Kotak \inst{2}
            \and
            K. Maguire \inst{1}
            \and
            A. Boyle \inst{3}
            }

	\institute{Astrophysics Research Centre, School of Mathematics and Physics, Queen's University Belfast, Belfast, BT7 1NN, UK \\ 				(\email{mmagee37@qub.ac.uk} \label{inst1})
	\and
	Tuorla Observatory, Department of Physics and Astronomy, FI-20014 University of Turku, Finland \label{inst2}
	\and
	Max-Planck-Institut f\"{u}r Astrophysik, Karl-Schwarzschild-Str. 1, 85748 Garching bei M\"{u}nchen, Germany \label{inst3}
		}

   \date{Received -	- -; accepted - - - }

 
  \abstract{Recent studies have argued that the progenitor system of type Iax supernovae must consist of a carbon-oxygen white dwarf accreting from a helium star companion. Based on existing explosion models invoking the pure deflagration of carbon-oxygen white dwarfs, we investigate the likelihood of producing spectral features due to helium in type Iax supernovae. From this scenario, we select those explosion models producing ejecta and $^{56}$Ni masses
  that are broadly consistent with those estimated for type Iax supernovae (0.014 -- 0.478~$M_{\odot}$ and $\sim0.003$ -- 0.183~$M_{\odot}$, respectively). To this end, we present a series of models of varying luminosities ($-18.4 \lesssim M_{\rm{V}} \lesssim -14.5$~mag) with helium abundances accounting for up to $\sim$36\% of the ejecta mass, and covering a range of epochs beginning a few days before B$-$band maximum to approximately two weeks after maximum. We find that the best opportunity for detecting \ion{He}{i} features is at near-infrared wavelengths, and in the post-maximum spectra of the fainter members of this class. We show that the optical spectrum of SN~2007J is potentially consistent with a large helium content (a few 10$^{-2}~M_{\odot}$), but argue that current models of accretion and material stripping from a companion struggle to produce compatible scenarios. We also investigate the presence of helium in all objects with near-infrared spectra. We show that SNe~2005hk, 2012Z, and 2015H contain either no helium or their helium abundances are constrained to much lower values ($\lesssim$10$^{-3}~M_{\odot}$). For the faint type Iax supernova, SN~2010ae, we tentatively identify a small helium abundance from its near-infrared spectrum. Our results demonstrate the differences in helium content among type Iax supernovae, perhaps pointing to different progenitor channels. Either SN~2007J is an outlier in terms of its progenitor system, or it is not a true member of the type Iax supernova class.}

   \maketitle
%

\section{Introduction}
\label{sect:intro}

Type Iax supernovae (SNe Iax) are a sub-class of type Ia supernovae that exhibit many unique properties \citep{02cx--orig, 02cx--late--spec, foley--13}. They do not follow the Phillip's relation \citep{phillips--99}, and are characterised by faint peak magnitudes ($-18.5 \lesssim$ $M_{\rm{V}}$ $\lesssim$ $-$14 mag) and low ejecta velocities ($\sim$2\,000 -- 9\,000~km~s$^{-1}$) \citep{02cx--orig,valenti--09,obs--08ha,obs--09ku,comp--obs--12z}. Their late-time spectra are unlike any other class of supernova and show significant diversity, containing both permitted and forbidden lines - the widths and strengths of which vary from object to object \citep{02cx--late--spec, foley--late--iax}. The extreme nature of SNe Iax has led to significant progress in the understanding of potential explosion scenarios. Proposed scenarios for SNe Iax include the pure deflagration of a carbon-oxygen white dwarf \citep{read--02cx--spectra,02cx--late--spec,phillips--07,foley--13}. In this scenario, subsonic burning of nuclear fuel (carbon and oxygen) is insufficient to fully unbind the star and produces well mixed ejecta, due to the turbulent propagation of the deflagration front. Alternatively, pulsational delayed detonations have also been suggested for at least the brightest SNe Iax \citep{02cx--orig,comp--obs--12z}, but may struggle to produce explosions with $^{56}$Ni masses as low as the faintest objects. 

\par

In addition to these explosion scenarios, the host environments of SNe Iax also provide constraints on the likely progenitor scenario. SNe Iax preferentially occur in young stellar environments and to date there has been no SN Iax discovered in an early type galaxy. The ages of stellar populations in the vicinity of SNe Iax \cite[$\lesssim$100~Myr;][]{08ha--prog,12z--prog} are consistent with the short delay time distribution predicted for helium-accreting carbon-oxygen white dwarf scenarios, while the lifetimes of hydrogen accreting systems appear too long to match what has been observed in SNe Iax - although this is based on only a few objects \citep{ruiter--09,wang--2009,wang--2010,wang--2013,wang--2014, liu--2015a}. The locations of SNe Iax within their host galaxies also appear correlated with ongoing star formation, similar to core collapse SNe, but unlike SNe Ia, which show no preference for star forming locations \citep{lyman--13,lyman--18}. This would further suggest progenitor scenarios involving relatively young stellar populations. Together, these arguments would appear to indicate that a helium donor to an accreting white dwarf is a viable candidate progenitor scenario for SNe Iax \citep{foley--13, liu--2015a}. 

\par

A review on the evolutionary channels that could lead to SNe Ia is given by \cite{wang--2012}. As discussed by these authors, there is a range of initial masses and periods that could lead to a viable helium donor channel. For systems containing stars of similar masses ($M_{1,{\rm i}}\sim5.5$ -- 6.5~$M_{\odot}$, $M_{2,{\rm i}}\sim5.0$ -- 6.0~$M_{\odot}$) and long initial periods ($P_{\rm i}\textgreater300$~days), the primary loses its hydrogen envelope during a common envelope phase, following unstable Roche lobe overflow. The hydrogen envelope of the secondary may also be lost at this phase, depending on when exactly the onset of RLOF from the primary begins. Alternatively, the secondary may lose its hydrogen envelope through an additional common envelope phase following subsequent evolution. The helium donor channel can also be formed from systems with much shorter initial periods ($P_{\rm i}\sim10$ -- 40~days) and more extreme mass ratios ($M_{1,{\rm i}}\sim5.0$ -- 8.0~$M_{\odot}$, $M_{2,{\rm i}}\sim2.0$ -- 6.5~$M_{\odot}$). The relative importance of these channels varies depending on the assumed common envelope ejection efficiency. \cite{wang--2009b} show how the helium donor channel can produce thermonuclear explosions with short delay times ($\sim45$ -- 140~Myr), making it a potential candidate for producing SNe Iax. 

\par

Further arguments in favour of helium donors include the detection of a progenitor to SN~2012Z (one of the brightest SNe Iax), as reported by \cite{12z--prog}. Based on similarities in colour and variability to the galactic helium nova V445~Puppis \citep{kato--2003}, they suggest that this object is the helium star companion to the white dwarf that exploded as SN~2012Z. The binary population synthesis calculations of \cite{liu--2015b} have also shown that this progenitor detection is consistent with a helium star donor. Although SN~2012Z is the only SN Iax with a potential progenitor detection, limits for other objects are also consistent with this scenario \citep{08ge--prog,14dt--prog}. 

\par

Finally, two objects have been potentially identified as SNe Iax with helium spectral features - SNe 2004cs and 2007J. SN~2007J was the first SN~Iax claimed to show helium features and has received classifications of SN~Iax \citep{07j--cbet--1, obs--08ha, foley--late--iax} and SN Ib/IIb \citep{07j--cbet--2, slow--ptf}. Based on spectroscopic similarities to SN~2007J, \cite{foley--13} argue that SN~2004cs is a SN~Iax showing helium features - despite also receiving a previous SN~IIb classification \citep{04cs--first}. The classification of both SN~2004cs and SN~2007J as SNe~Iax is disputed by \cite{slow--ptf}, who argue that neither object is a true member of the class, and therefore the presence of helium in their spectra should not be used to constrain the entire class - a claim rejected by \cite{foley--late--iax}. It is therefore clear that the question of whether SNe Iax contain helium at all, and if so, quantifying this amount is one with important consequences for the class and their likely progenitor scenarios. 

\par

Motivated by the above, we aim to investigate in this study how much helium could be contained within the ejecta of SNe Iax, whether these quantities would result in spectral features that would be detectable in the optical and near-IR spectra of SNe Iax, and how this might constrain various progenitor channels. We discuss the models used in this study in Sect.~\ref{sect:models},
and present synthetic spectra based on these in Sect.~\ref{sect:model_spectra}. In Sect.~\ref{sect:comparisons} we compare these synthetic spectra to existing observations of SNe Iax. We focus our analysis on SN~2007J, as it has been claimed to show features due to helium and has publicly available spectra, and all existing SNe Iax with infrared spectra, due to the strong \ion{He}{i}~$\lambda$10\,830 transition and the relatively clean nature of this region of the spectra. In Sect.~\ref{sect:discussion} we discuss potential sources of helium, and conclude in Sect.~\ref{sect:conclusions}.

%

\section{Models}
\label{sect:models}

\begin{table}
\centering
\caption{Explosion model properties}\tabularnewline
\label{tab:model-props}\tabularnewline
\resizebox{\columnwidth}{!}{
\begin{tabular}{lccccc}\hline
\hline
Model &	Ejecta mass & $^{56}$Ni mass & Kinetic energy & Peak $M_{B}$ & $\Delta m_{15}(B)$ \tabularnewline
 & $M_{\odot}$ & $M_{\odot}$ & 10$^{50}$~erg & (mag) & (mag) \tabularnewline
\hline
N1def & 0.084 & 0.035 & 0.149 & $-16.55$ & 2.15 \tabularnewline
N3def & 0.195 & 0.073 & 0.439 & $-17.55$ & 1.91 \tabularnewline
N5def & 0.372 & 0.158 & 1.350 & $-17.85$ & 1.69 \tabularnewline
N10def & 0.478 & 0.183 & 1.950 & $-17.95$ & 1.68 \tabularnewline
N5def-hybrid & 0.014 & 0.003 & 0.018 & $-14.12$ & 2.24 \tabularnewline
\hline
\end{tabular}
}
\tablefoot{Properties of the explosion models used as reference in this study. Values are taken from \cite{fink-2014} and \cite{kromer-15}.}
\end{table}

\begin{table}
\centering
\caption{TARDIS model parameters and properties}\tabularnewline
\label{tab:model-params}\tabularnewline
\resizebox{\columnwidth}{!}{
\begin{tabular}{ccccc}\hline
\hline
\multicolumn{3}{c}{TARDIS input parameters} \vline & \multicolumn{2}{c}{Derived model properties} \tabularnewline
\hline
Time since  & Luminosity & Inner boundary  & Phase & Inner boundary  \tabularnewline
explosion (days) & ($\log$~L$_{\odot}$) & velocity (km~s$^{-1}$) & (days) & temperature (K)  \tabularnewline
\hline
\hline
\multicolumn{5}{c}{N1def} \tabularnewline
\hline
$\phn$2.6    & 8.17  	 & $\phn$7\,500     &  $\phn-$5    &  19\,200           \tabularnewline
$\phn$7.7    & 8.48       & $\phn$6\,500    &  $\phn$+0    &  11\,300          \tabularnewline
12.4         & 8.35       & $\phn$6\,100    &  $\phn$+5    &  $\phn$8\,000     \tabularnewline
22.4         & 7.94       & $\phn$4\,100    &  +15         &  $\phn$5\,600     \tabularnewline
\hline
\multicolumn{5}{c}{N3def} \tabularnewline
\hline
$\phn$5.0   & 8.59  	 & $\phn$9\,700    &  $\phn-$5     &    11\,500       \tabularnewline
10.0        & 8.73       & $\phn$7\,100    &  $\phn$+0     &    11\,000        \tabularnewline
14.9        & 8.65       & $\phn$6\,800    &  $\phn$+5     &    $\phn$8\,200   \tabularnewline
25.0        & 8.30       & $\phn$4\,700    &  +15          &    $\phn$6\,400   \tabularnewline
\hline
\multicolumn{5}{c}{N5def} \tabularnewline
\hline
$\phn$6.2   & 8.86  	 & 10\,300         &   $\phn-$5     &  13\,400        \tabularnewline
11.1        & 9.01       & $\phn$7\,800    &   $\phn$+0     &  13\,500        \tabularnewline
16.1        & 8.94       & $\phn$6\,900    &   $\phn$+5     &  10\,000        \tabularnewline
26.0        & 8.66       & $\phn$5\,900    &   +15          &  $\phn$7\,000   \tabularnewline
\hline
\multicolumn{5}{c}{N10def} \tabularnewline
\hline
$\phn$6.2   & 8.89  	  & 10\,300       & $\phn-$5     &  14\,700         \tabularnewline
11.1        & 9.06       & $\phn$8\,400   & $\phn$+0     &  13\,600         \tabularnewline
16.1        & 9.01       & $\phn$7\,100   & $\phn$+5     &  10\,700           \tabularnewline
26.0        & 8.74       & $\phn$6\,400   & +15          &  $\phn$7\,200      \tabularnewline
\hline
\multicolumn{5}{c}{N5def-hybrid} \tabularnewline
\hline
$\phn$1.8   & 7.45   	 & $\phn$5\,000    & $\phn-$2     &  18\,800        \tabularnewline
$\phn$3.8   & 7.60       & $\phn$4\,200    & $\phn$+0     &  13\,700        \tabularnewline
$\phn$8.8   & 7.37       & $\phn$3\,000    & $\phn$+5     &  $\phn$8\,000   \tabularnewline
$\phn$18.8  & 6.78       & $\phn\phn$600   & +15          &  $\phn$8\,900   \tabularnewline
\hline
\end{tabular}
}
\tablefoot{Input parameters used for our TARDIS models. Phases are given relative to the time of bolometric maximum.}
\end{table}

Here, we present a series of simulations with varying helium abundances calculated using TARDIS, a one-dimensional Monte Carlo radiative transfer code \citep{tardis, tardis_v2}. We use these simulations to investigate the effects of helium on the synthetic spectra across a range of epochs and peak luminosities. 

\par

Our simulations are based on the multidimensional explosion simulations of \cite{fink-2014} and \cite{kromer-15}. These models invoke the pure deflagration of Chandrasekhar-mass carbon-oxygen \citep{fink-2014} and hybrid \citep{kromer-15} white dwarfs, with ZAMS solar metallicities. Within the \cite{fink-2014} model sequence, the strength of the explosion (and hence the amount of $^{56}$Ni produced) is controlled by the number of ignition sparks used to ignite the deflagration. Models within this sequence are named corresponding to the number of sparks (i.e. N5def was ignited with five sparks) and generally form a trend of increasing luminosity. The N5def-hybrid model does not fit this trend. Deflagrations of hybrid white dwarfs are most suitable for the faintest SNe Iax, such as SN~2008ha \citep{kromer-15}. In this scenario, the carbon-oxygen core is surrounded by an oxygen-neon mantle. The presence of the mantle halts the propagation of the deflagration front and therefore only a small amount of $^{56}$Ni is produced ($\sim3 \times 10^{-3}~M_{\odot}$). It is unclear however, whether such pure deflagrations in hybrid white dwarfs are realised in nature \citep{brooks--17}. In Table~\ref{tab:model-props} we list the properties of the explosion models used as references in this study \citep{fink-2014, kromer-15}.

\par

We select the N1def, N3def, N5def, N10def, and N5def-hybrid models as they lie in the appropriate luminosity range for SNe Iax \cite[$-18$ $\lesssim M_{\rm{B}}$ $\lesssim$ $-$14 mag;][]{fink-2014,kromer-15}. The N3def, N5def, and N5def-hybrid models have previously been the subject of in-depth analyses and comparisons to existing observations of SNe Iax \citep{kromer-13, kromer-15, 15h}. Radiative transfer simulations show these models are able to broadly reproduce the luminosity and spectral features of SNe 2015H, 2005hk, and 2008ha. There are however, discrepancies between the models and the data: the models appear to systematically eject a lower mass than suggested by observations of SNe Iax, and hence also show faster evolving light curves. Therefore, we stress that the series of models presented in this study is intended to explore the effects of helium on the model spectra for a variety of peak luminosities, and not to provide fits to individual objects. Our goal is to investigate whether the increased abundance of helium in the ejecta improves the agreement with observations, compared to models that contain little or no helium.

\par

For all of the TARDIS models presented in this study, we use the helium treatment described by \cite{boyle--17}, who applied this to an investigation of spectral features produced by helium shell detonation models. As discussed by \cite{hachinger--12}, who studied stripped envelope core collapse SNe, the unique atomic structure of helium causes strong departures from local thermodynamic equilibrium (LTE) for the conditions typical of SN ejecta. The ionisation state of helium is strongly affected by non-thermal electrons produced by interactions with $\gamma$-rays from the decay of $^{56}$Ni and $^{56}$Co. This results from the fact that the \ion{He}{i} ground state is separated from all other levels by a large energy gap and the lowest excited levels of \ion{He}{i} are metastable, causing them to be more strongly coupled to the \ion{He}{ii} population than the \ion{He}{i} ground state. Therefore, in the approximation of \cite{boyle--17}, the ion and excited level populations of helium are calculated relative to the population of the \ion{He}{ii} ground state, while the \ion{He}{i} ground state population is assumed to be negligible. This approximation is valid provided helium exists predominantly in a singly-ionised state. As shown by \cite{dessart--15} for their helium shell detonation models, helium remains ionised for weeks after maximum light. The models presented in this study cover a similar range in temperatures to those of \cite{dessart--15}, however the ejecta structure differs significantly. The aforementioned study used a helium shell detonation model that produced an ejecta mass of 0.2~$M_{\odot}$ and set the maximum velocity to $\sim$30\,000~km~s$^{-1}$, whereas the models used in this work cover a broad range of ejecta masses ($\sim$0.014 -- 0.478~$M_{\odot}$) and maximum velocities ($\sim$6\,000 -- 13\,000~km~s$^{-1}$). The ejecta densities of our models are generally higher than the \cite{dessart--15} model, which likely results in a decrease of the ionisation fraction. However, our models contain significantly more $^{56}$Ni, which could provide a greater source of ionising particles. The extent to which either of these properties dominates or produces a noticeably different ionisation fraction relative to the \cite{dessart--15} model is unclear. 

\par

In addition, the degree of ionisation will also affect the fraction of $^{56}$Ni decay energy contributing towards heating the gas, as opposed to ionising it. As shown by \cite{kozam-92}, the fraction of energy that heats thermal electrons increases with an increasing electron fraction, in other words this results from a more highly ionised gas. A detailed treatment of non-thermal processes is necessary to test the validity of this assumption for the specific circumstances of the explosions explored as part of this work, but this is beyond the scope of the simple exploratory work presented here. We also note that \cite{boyle--17} show how the approximations used in this study produce helium level populations comparable to those calculated from a full non-LTE system of statistical equilibrium equations, using the statistical equilibrium solver of \cite{hachinger--12}.

\par

\begin{figure}
\centering
\includegraphics[width=\columnwidth]{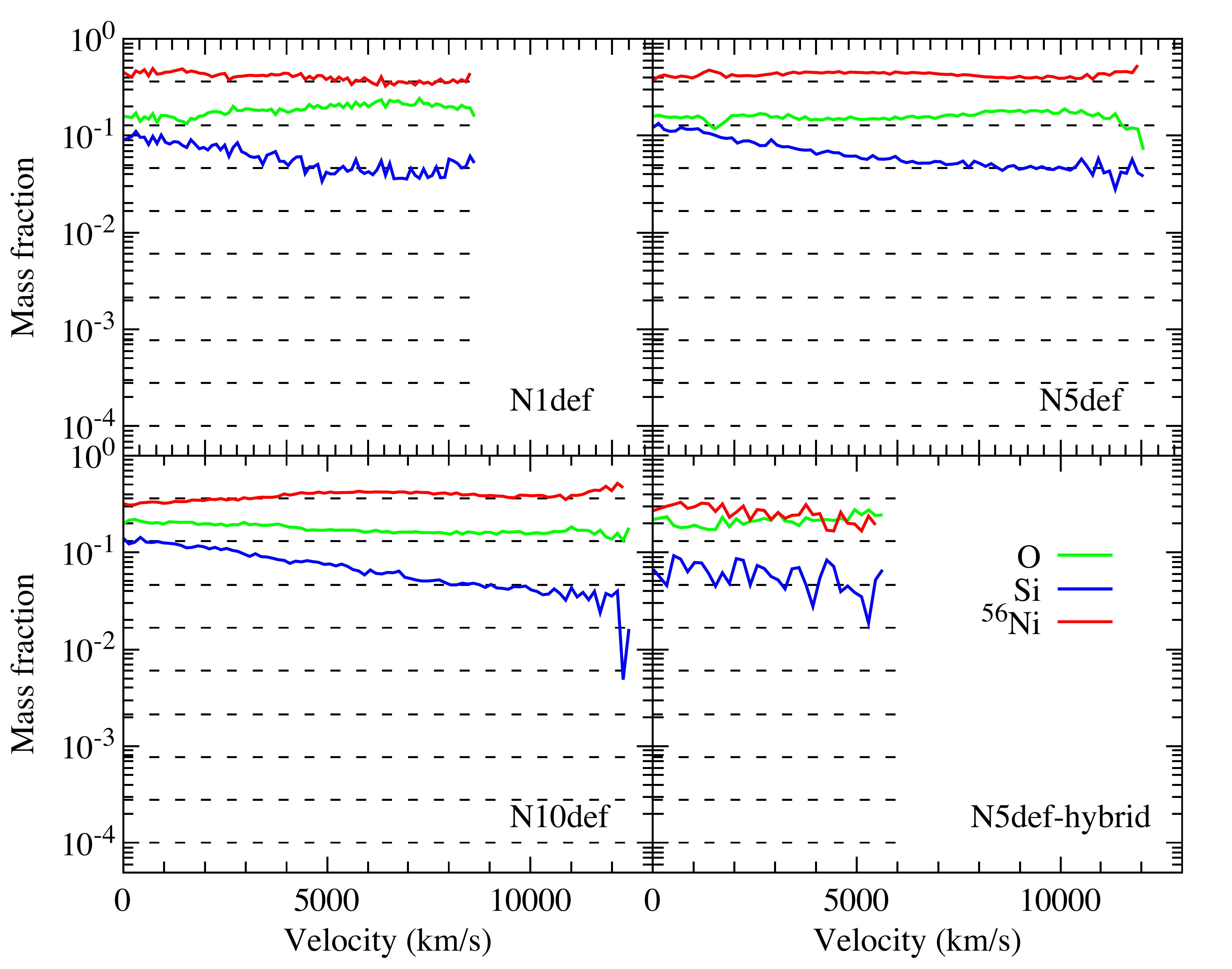}
\caption{Illustrative compositions at maximum light for some of models used in this study. With the exception of helium, mass fractions of all elements are taken directly from the \cite{fink-2014} and \cite{kromer-15} deflagration models. Helium abundances are shown as dashed horizontal lines. As these pure deflagration models predict fully mixed ejecta, we have chosen to maintain uniform helium abundances throughout.}
\label{fig:composition}
\centering
\end{figure}

\begin{figure*}[!h]
\centering
\includegraphics[width=\textwidth]{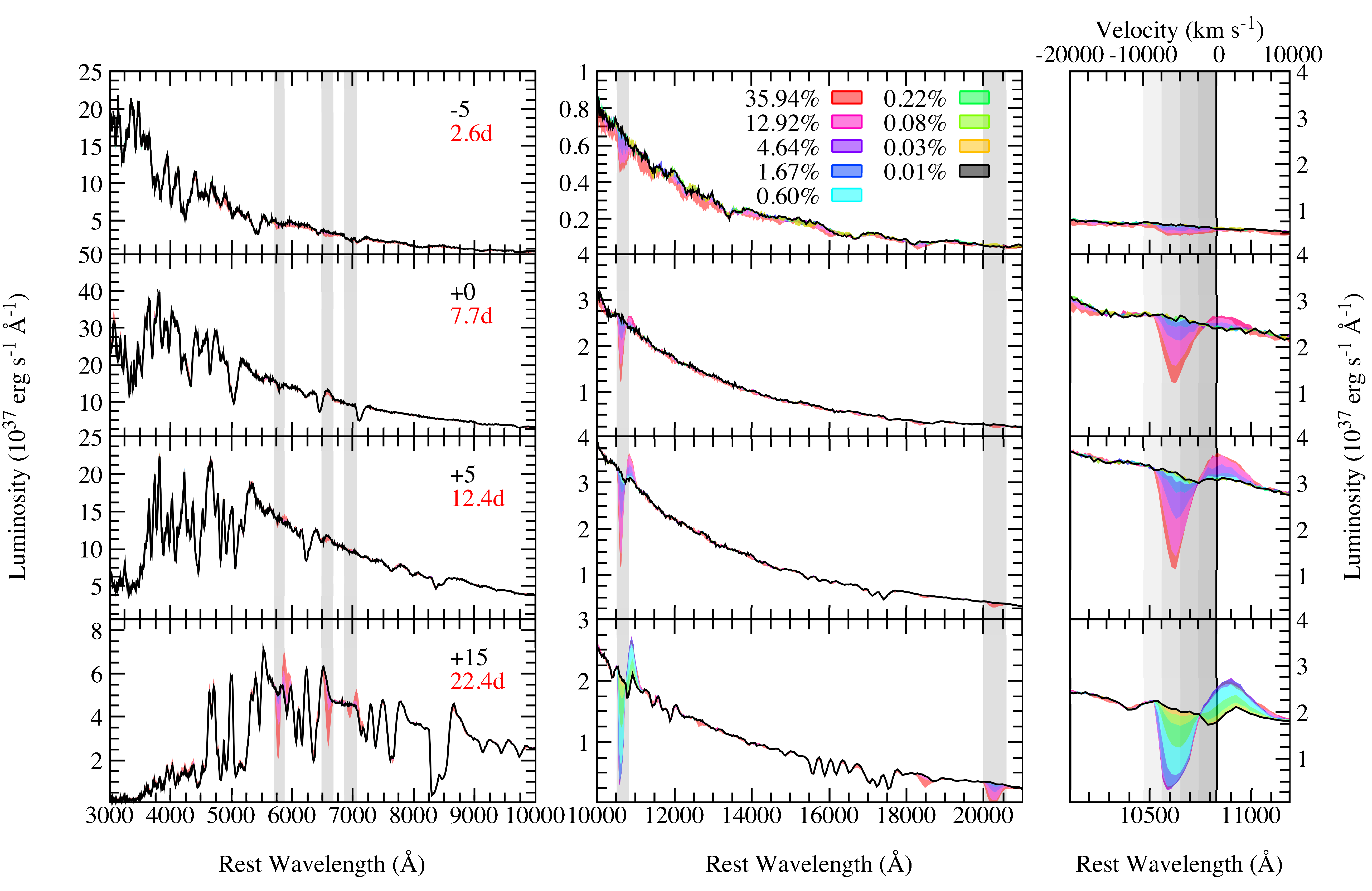}
\caption{N1def model spectra at various epochs, with varying helium abundances given as percentages of the ejecta mass. Shaded regions show the difference produced in the spectra by varying the helium abundance. Phases relative to bolometric maximum light are given in black, while times since explosion are given in red. {\it Left:} Grey shaded regions represent \ion{He}{i}~$\lambda$5\,876, $\lambda$6\,678, $\lambda$7\,065, $\lambda$10\,830, and $\lambda$20\,587 at rest and blue-shifted by the maximum velocity of the model ($\sim$8\,800~km~s$^{-1}$). {\it Right:} Zoomed in regions surrounding \ion{He}{i}~$\lambda$10\,830. Each shaded region corresponds to a blue-shift of 2\,500~km~s$^{-1}$. 
}
\label{fig:n1def_varying_helium}
\centering
\end{figure*}

TARDIS requires a number of parameters as input for each simulation. These are: the time since explosion, the luminosity of the supernova, the position of the inner boundary (i.e., the photosphere), and the composition and density of the ejecta. The input parameters we use for each model are listed in Table~\ref{tab:model-params}. We use the bolometric light curves presented by \cite{fink-2014} and \cite{kromer-15} to determine the luminosity and time since explosion for each TARDIS simulation. The inner boundary defines the position of the photosphere, which is required for TARDIS simulations and separates the optically thick and thin regions. Due to this assumption, we limit our investigation to phases around maximum light: $-5$ (the $\sim$4 day rise time of the N5def-hybrid model necessitates a different phase of $-2$), $+0$, $+5$, and $+15$ days. At later epochs, this assumption is unlikely to be valid. We determine values for the inner boundaries based on the velocities observed in the spectra of each explosion model. For each explosion model and epoch, we take an average velocity from at least two optical iron lines that are fit with a Gaussian profile.
In an effort to quantify the effect of our choice of inner boundary, we also calculate synthetic spectra for models using inner boundary velocities of $\pm$500~km~s$^{-1}$ and $\pm$1\,000~km~s$^{-1}$ for N3def, N5def and N10def, and $\pm$250~km~s$^{-1}$ and $\pm$500~km~s$^{-1}$ for N1def and N5def-hybrid. We find that our conclusions are unaffected by the choice of inner boundary velocity within these velocity uncertainties. 

\par

Finally, TARDIS requires the ejecta structure to be defined. The density profile for each TARDIS model is taken as the angle-averaged density of the multidimensional explosion simulations. With the exception of helium, the elemental abundances are also taken as angle-averages of the abundance profiles from the explosion simulations. We include elements up to Zn and typically ions up to eight times ionised. Given the temperatures covered in our models, it is unlikely that a significant population of more highly ionised species is present.
As we wish to test the effects of helium on the spectra, we arbitrarily choose helium mass fractions where $-4 \lesssim \log$~X$_{\rm{He}} \lesssim -0.4$, while keeping the total ejecta mass constant and scaling down the relative mass fractions of every other element. The pure deflagration models of \cite{fink-2014} and \cite{kromer-15} predict that the ejecta is nearly fully mixed on a macroscopic scale (current simulations are not able to resolve microscopic mixing, however this may influence the degree of ionisation) with a uniform composition, we have therefore opted to use a constant helium mass fraction for each model. The mass fractions of some of the most abundant elements are shown in Fig.~\ref{fig:composition} for some of the models studied. As shown in Fig.~\ref{fig:composition}, the $^{56}$Ni mass fraction is nearly uniform throughout the model, and therefore always co-exists with helium. Finally, for our simulations we use the ``macro-atom'' method for interactions between photons and elements within the ejecta. This is the most sophisticated line interaction mode within TARDIS and allows atoms to undergo multiple internal transitions after absorbing a photon. Further details are given in \cite{tardis}, but see also \cite{lucy-02, lucy-03}

\begin{figure*}
\centering
\includegraphics[width=\textwidth]{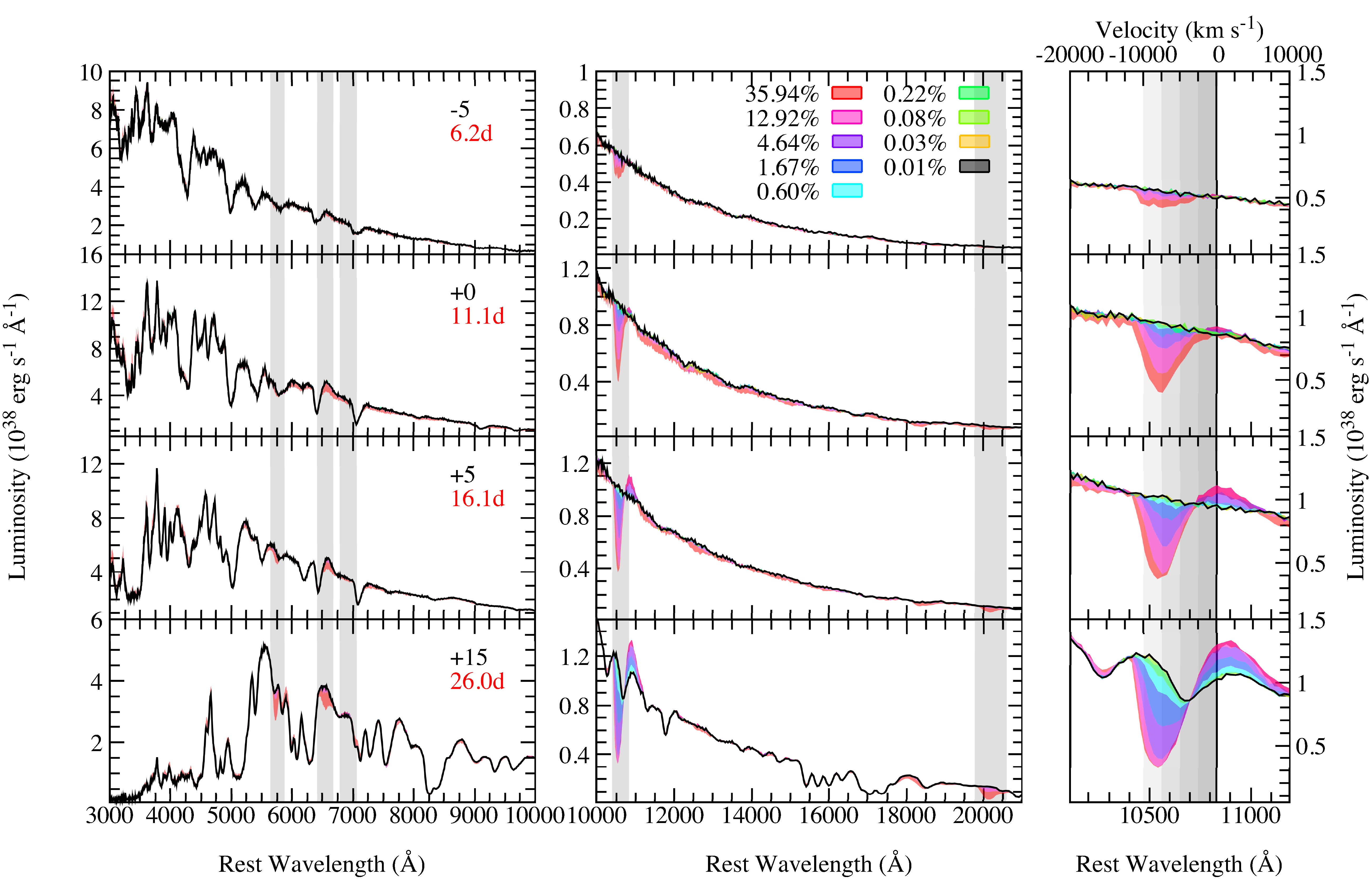}
\caption{N5def model spectra at various epochs, with varying helium abundances given as percentages of the ejecta mass. Shaded regions show the difference produced in the spectra by varying the helium abundance. Phases relative to bolometric maximum light are given in black, while times since explosion are given in red. {\it Left:} Grey shaded regions represent \ion{He}{i}~$\lambda$5\,876, $\lambda$6\,678, $\lambda$7\,065, $\lambda$10\,830, and $\lambda$20\,587 at rest and blue-shifted by the maximum velocity of the model ($\sim$12\,300~km~s$^{-1}$). {\it Right:} Zoomed in regions surrounding \ion{He}{i}~$\lambda$10\,830. Each shaded region corresponds to a blue-shift of 2\,500~km~s$^{-1}$.
}
\label{fig:n5def_varying_helium}
\centering
\end{figure*}

%

\section{Model spectra}
\label{sect:model_spectra}

In this section we discuss general trends among the models, while Sect.~\ref{sect:comparisons} focuses on comparing synthetic spectra to observations of SNe Iax. In Figs.~\ref{fig:n1def_varying_helium} and \ref{fig:n5def_varying_helium} we show the spectra generated for our N1def and N5def models, respectively. These models are representative of the faintest and brightest SNe Iax explosions. Similar figures for our N3def, N10def, and N5def-hybrid models are shown in Figs.~\ref{fig:n3def_varying_helium},~\ref{fig:n10def_varying_helium}, and \ref{fig:n5def_hybrid_varying_helium}, respectively. Each model is shown at four epochs and with varying helium abundances. 

\par

Our models show that early time spectra are not ideal for observing helium, regardless of the underlying explosion strength. Only those models with large helium abundances, of the order of tens of percent, show the \ion{He}{i}~$\lambda$10\,830 feature before maximum light -- although even in these cases, this feature is generally quite weak. Moving to later times, the helium features continue to increase in strength. This is likely due to the decrease in radiative temperature throughout the model, and subsequent increase in the \ion{He}{i} number density. We also note that the appearance of the \ion{He}{i} features over time should be affected by changes in the $\gamma$-ray mean free path, and hence in the \ion{He}{ii} population. As discussed in Sect.~\ref{sect:models}, we assume in this work that helium is predominantly singly ionised. Full NLTE simulations are necessary to test what effect this will have on the population of helium that is singly ionised.

\par

By maximum light, the larger helium abundance models now show much stronger \ion{He}{i}~$\lambda$10\,830, and this feature has also started to emerge for the lower abundance models. For example, the N5def model containing 36\% helium shows an approximately five-fold increase in the strength of \ion{He}{i}~$\lambda$10\,830 between the pre-maximum and maximum light spectra. This trend persists through to $+15$\,d post-maximum, by which time the strength of the  \ion{He}{i}~$\lambda$10\,830 feature relative to the continuum in the N5def model containing 36\% helium has increased by a further factor of approximately three relative to maximum light. 

\par

As expected, our models show that the \ion{He}{i} features in the near-IR region are much more prominent than optical features. The optical features themselves are much more difficult to excite and are only produced in models with large helium abundances ($\gtrsim$10\%) at later times. Regardless of peak luminosity, phase, or helium mass fraction, features due to \ion{He}{i}~$\lambda$5\,876, $\lambda$6\,678, and $\lambda$7\,065 are not produced without an accompanying strong \ion{He}{i}~$\lambda$10\,830 feature. 

\par

We find that lower luminosity models typically show stronger helium features relative to their higher luminosity counterparts. Again, this is likely due to the lower radiative temperature and higher \ion{He}{i} population density in these models.
For example, Fig.~\ref{fig:n1def_varying_helium} shows that at $+15$ days post-maximum, the N1def models containing 0.08\% and 0.03\% helium can be easily distinguished between based on the strengths of their \ion{He}{i}~$\lambda$10\,830 features. This is not the case for the N5def models, where models with helium abundances $\lesssim$0.22\% do not show strong helium features, and hence it is not possible to distinguish between them (see Fig.~\ref{fig:n5def_varying_helium}).

\par

In summary, our models suggest that, as long as helium remains predominantly singly ionised, helium features should be strongest at later times and in fainter SNe Iax. Optical features can be produced in models with sufficiently large helium abundances, but are always accompanied by \ion{He}{i}~$\lambda$10\,830 as the strongest feature.

%

\section{Comparisons with observed spectra of SNe Iax}
\label{sect:comparisons}
In this section we compare our synthetic spectra to individual SNe. We select all SNe Iax with infrared spectra and SN~2007J, which has been argued to show strong optical helium features (see Sect.~\ref{sect:intro}). All observed spectra have been corrected for redshift, galactic and host extinction, where appropriate.

\subsection{SN~2007J}

\begin{figure*}[!t]
\centering
\includegraphics[width=\textwidth]{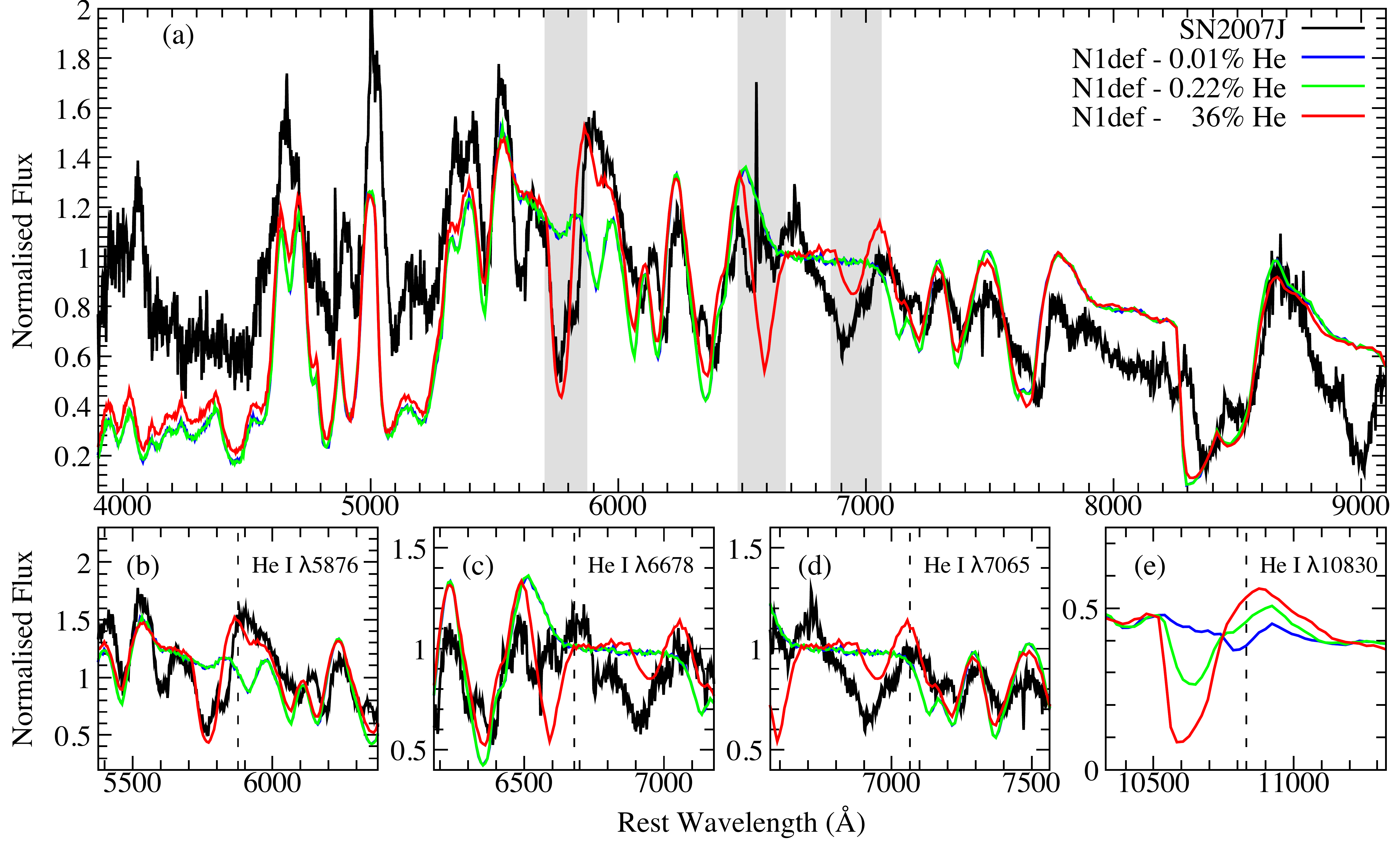}
\caption{Comparison of SN~2007J and our N1def models with 0.01\%, 0.22\%, and 36\% helium abundances. Spectra have been normalised to the median flux between 5\,000\,\AA\, -- 7\,000\,\AA. The spectrum of SN~2007J is estimated to lie between $+6$ -- $+46$ days after $V$-band maximum and has been corrected for galactic extinction only. Our N1def model is shown at $+15$ days after bolometric maximum light. (a): Shaded regions represent \ion{He}{i}~$\lambda$5\,876, $\lambda$6\,678, $\lambda$7\,065, and $\lambda$10\,830 at rest and blue-shifted by the maximum velocity of the model ($\sim$8\,800~km~s$^{-1}$). (b), (c), (d), and (e): Zoom-ins of the regions surrounding \ion{He}{i}~$\lambda$5\,876, $\lambda$6\,678, $\lambda$7\,065, and  $\lambda$10\,830, respectively. In panel (c) we show the host subtracted spectrum. Rest wavelengths are denoted by a vertical dashed line. }
\label{fig:07J_helium_comparison}
\centering
\end{figure*}

In Fig.~\ref{fig:07J_helium_comparison}, we show a comparison between SN~2007J \citep{foley--13} and our N1def models with 0.01\%, 0.22\%, and 36\% helium abundances. The peak absolute $V$-band magnitude of SN~2007J is not well constrained, but was $\lesssim-$15.4~mag, while the N1def model peaked at $M_{\rm{V}} = -16.8$~mag. Although the epoch of maximum for SN~2007J is highly uncertain \cite[the spectrum presented in Fig.~\ref{fig:07J_helium_comparison} is estimated to lie between +6 and +46 days;][]{foley--13}, we find that our N1def synthetic spectra at +15 days show good agreement. Overall, the shape of the spectrum deviates slightly from what is observed in SN~2007J, with the model producing less flux than is observed at $\lesssim$5\,000\,\AA. The N1def synthetic spectra are able to reproduce many of the features in terms of their line strengths, velocities, and shapes, however notable exceptions include the strong absorption features at $\sim$5\,800\,\AA, $\sim$6\,900\,\AA, and $\sim$9\,000\,\AA.

\par

As shown in Fig.~\ref{fig:07J_helium_comparison}, increasing the helium abundance to 0.22\% has little effect on the optical spectrum, but does produce a strong feature due to \ion{He}{i}~$\lambda$10\,830. Unfortunately no infrared spectra are available for SN~2007J to compare against this prediction. From Fig.~\ref{fig:07J_helium_comparison}(b), (c), and (d) it is clear that this increase in helium abundance, by more than an order of magnitude, still fails to produce any noticeable signs of the \ion{He}{i} optical features. Increasing the helium abundance further, to 36\% of the ejecta mass, provides the best match to the optical spectrum of SN~2007J from the models explored here. As shown in Fig.~\ref{fig:07J_helium_comparison}(b), the increased abundance of helium produces a strong absorption feature due to \ion{He}{i}~$\lambda$5\,876 that is well matched in SN~2007J. Our model with 0.01\% helium abundance does produce a feature due to \ion{Fe}{ii} at $\sim$5\,750\,\AA, however this feature is much weaker than that observed in SN~2007J. In addition, the emission peak from the \ion{He}{i}~$\lambda$5\,876 P-Cygni profile fills in the flux that was removed due to the \ion{Fe}{ii} feature at $\sim$5\,900\,\AA\, and provides overall better agreement with the red wing. It is conceivable that the ionisation state of the ejecta could be altered to bring these \ion{Fe}{ii} features into better agreement with SN~2007J, however such a change would likely have significant consequences for the rest of the optical spectrum and worsen the match with SN~2007J.

\par

The \ion{He}{i}~$\lambda$6\,678 feature is produced in the model, but is not clearly visible in SN~2007J (Fig.~\ref{fig:07J_helium_comparison}(c)). The spectrum of SN~2007J is clearly contaminated by H$\alpha$ and \ion{N}{ii} emission from the host galaxy, which makes it difficult to discriminate exactly between features in this region. In an effort to better discern the features in this region, we extracted a spectrum of the host galaxy offset from SN~2007J. We arbitrarily scale this spectrum to approximately match the flux of the host contamination in SN~2007J and then subtract it. In Fig.~\ref{fig:07J_helium_comparison}(c) we show the spectrum of SN~2007J after subtraction of the host galaxy.
Removing the host contamination we find that the spectrum of SN~2007J could contain two relatively narrow features centred around $\sim6\,540$ and $\sim$6\,630~$\AA$ (see Fig.~\ref{fig:07J_helium_comparison}(c)). The redder of these two features could be \ion{He}{i}~$\lambda$6\,678, although with a slightly lower velocity and strength than predicted by our model. We note however, that the velocity offset between the model and observations for this feature ($\sim$1\,500~km~s$^{-1}$) is similar to the offset between the observed and model \ion{Fe}{ii} features centred around $\sim6\,100$ and $\sim$6\,200~$\AA$ (see Fig.~\ref{fig:07J_helium_comparison}(b)). We therefore deem it likely that \ion{He}{i}~$\lambda$6\,678 is present in the spectrum of SN~2007J to some degree. The identification of the bluer feature (at $\sim6\,530$~$\AA$) is not clear, as this region is dominated by the strong \ion{He}{i}~$\lambda$6\,678 feature of our model. A potential identification could be broad (relative to the host lines), low velocity H$\alpha$ originating from the supernova, rather than the host. This feature does not clearly appear in later spectra. We note however, that the level of host contamination increases with time, as the supernova flux decreases. Therefore it is difficult to make a conclusive identification and we consider the H$\alpha$ identification as speculative only.

\par

 The model also produces an absorption feature due to \ion{He}{i}~$\lambda$7\,065 (Fig.~\ref{fig:07J_helium_comparison}(d)) that is approximately coincident with a feature in SN~2007J. This feature could be the result of a blend between \ion{He}{i}~$\lambda$7\,065 and an additional line that is not produced in our model - the spectrum of SN~2007J appears to show a small red shoulder. If this is not due to a blend, then the model feature is clearly much weaker than in SN~2007J. We note however, that the \ion{He}{i}~$\lambda$5\,876 and $\lambda$7\,065 transitions share the same lower level (1s2p $^3$P) and therefore it is unlikely that the strength of one feature could be dramatically altered without affecting the other.

\begin{figure*}[!t]
\centering
\includegraphics[width=\textwidth]{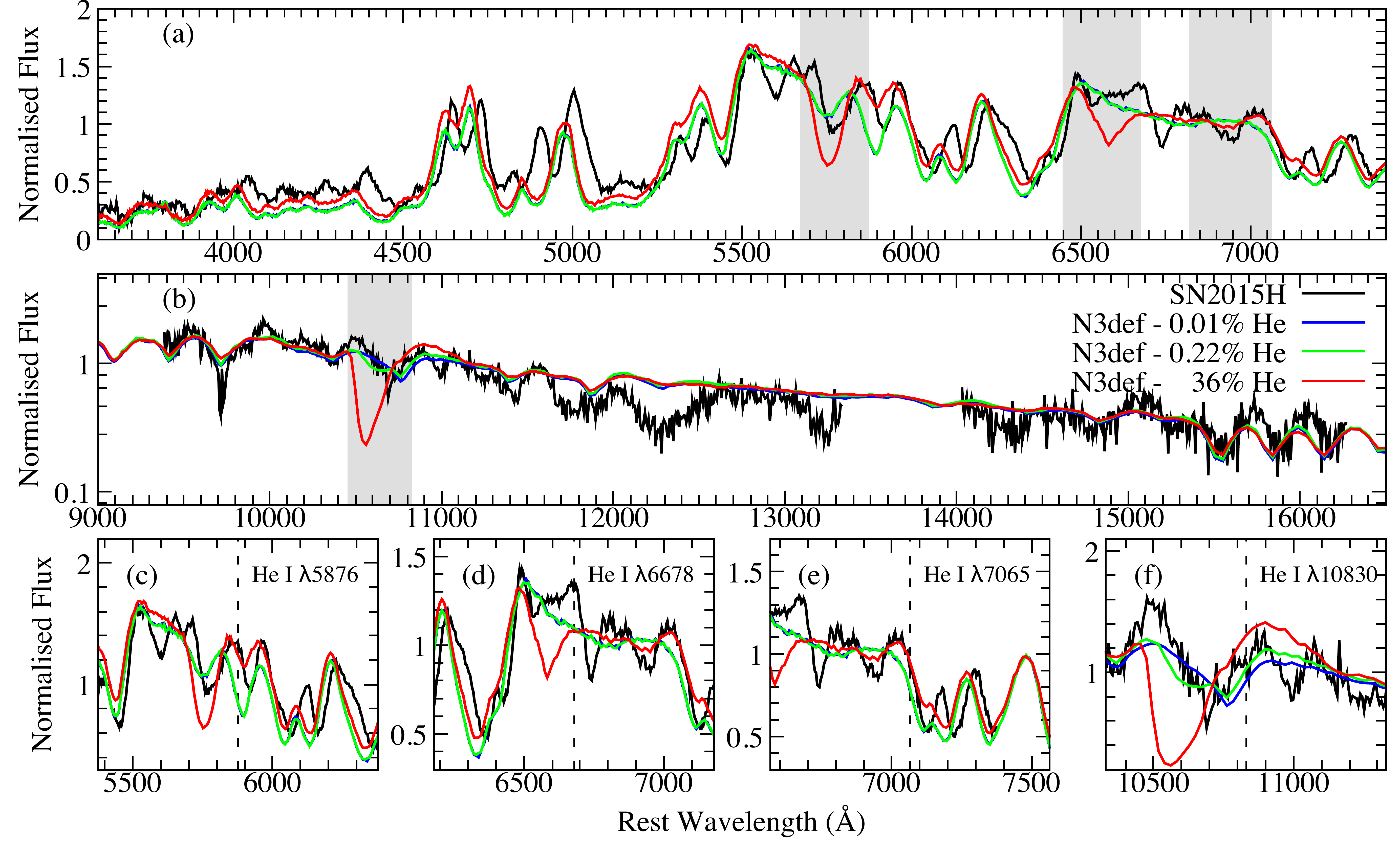}
\caption{Comparison of SN~2015H and our N3def models with 0.01\%, 0.22\%, and 36\% helium abundances. Optical and infrared spectra have been normalised to the median flux from 5\,000\,\AA\, -- 7\,000\,\AA\, and 10\,000\,\AA\, -- 12\,000\,\AA, respectively. The spectra of SN~2015H are approximately $+16$ days after $r$-band maximum and have been corrected for galactic extinction only. Our N3def model is shown at $+15$ days after bolometric maximum light. (a) and (b): Shaded regions represent \ion{He}{i}~$\lambda$5\,876, $\lambda$6\,678, $\lambda$7\,065 and $\lambda$10\,830 at rest and blue-shifted by the maximum velocity of the model ($\sim$10\,400~km~s$^{-1}$). (c), (d), (e) and (f): Zoom-ins of the regions surrounding \ion{He}{i}~$\lambda$5\,876, $\lambda$6\,678,  $\lambda$7\,065, and  $\lambda$10\,830, respectively. Rest wavelengths are denoted by a vertical dashed line.
}
\label{fig:15H_helium_comparison}
\centering
\end{figure*}

\par

The features of SN~2007J reported as being due to helium by \cite{foley--13} were observed to increase in strength over time (including the potential \ion{He}{i}~$\lambda$6\,678 feature) between spectra with phases ranging from approximately $+6$ -- $+46$ days to $\sim$+62 -- $+102$ days post maximum light \citep{foley--13}; an increase in \ion{He}{i} line strengths is also apparent in our model sequence between maximum light and +15\,d (see Fig.~\ref{fig:n1def_varying_helium}). Overall our N1def model with a helium mass fraction of 36\% shows remarkably good agreement with SN~2007J, suggesting there could be a significant helium mass fraction, on the order of tens of percent, equivalent to a few hundredths of a solar mass, in the ejecta. Section~\ref{sect:discussion} discusses the implications of this result for progenitor scenarios.

%

\subsection{SN~2015H, SN~2005hk, and SN~2012Z}

\begin{figure*}[!t]
\centering
\includegraphics[width=\textwidth]{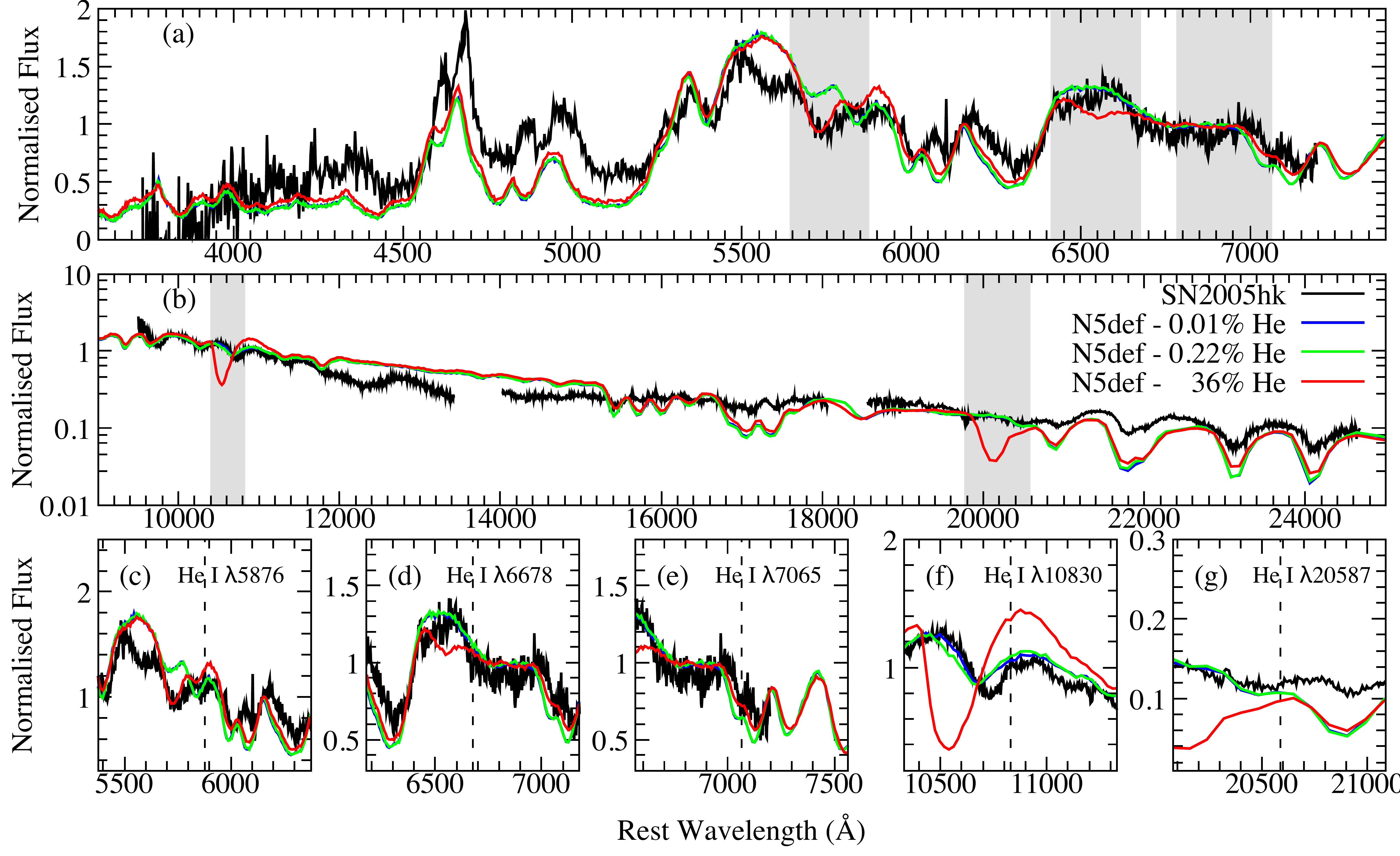}
\caption{Comparison of SN~2005hk and our N5def models with 0.01\%, 0.22\%, and 36\% helium abundances. Optical and infrared spectra have been normalised to the median flux from 5\,000\,\AA\, -- 7\,000\,\AA\, and 10\,000\,\AA\, -- 12\,000\,\AA, respectively. The optical and infrared spectra of SN~2005hk are approximately $+13$ days after $B$-band maximum and have been corrected for galactic extinction only. Our N5def model is shown at $+15$ days after bolometric maximum light. (a) and (b): Shaded regions represent \ion{He}{i}~$\lambda$5\,876, $\lambda$6\,678, $\lambda$7\,065, $\lambda$10\,830, and $\lambda$20\,587 at rest and blue-shifted by the maximum velocity of the model ($\sim$12\,300~km~s$^{-1}$). (c), (d), (e), (f), and (g): Zoom-ins of the regions surrounding \ion{He}{i}~$\lambda$5\,876, $\lambda$6\,678,  $\lambda$7\,065, $\lambda$10\,830, and $\lambda$20\,587, respectively. Rest wavelengths are denoted by a vertical dashed line.}
\label{fig:05hk_helium_comparison}
\centering
\end{figure*}

\begin{figure*}[!t]
\centering
\includegraphics[width=\textwidth]{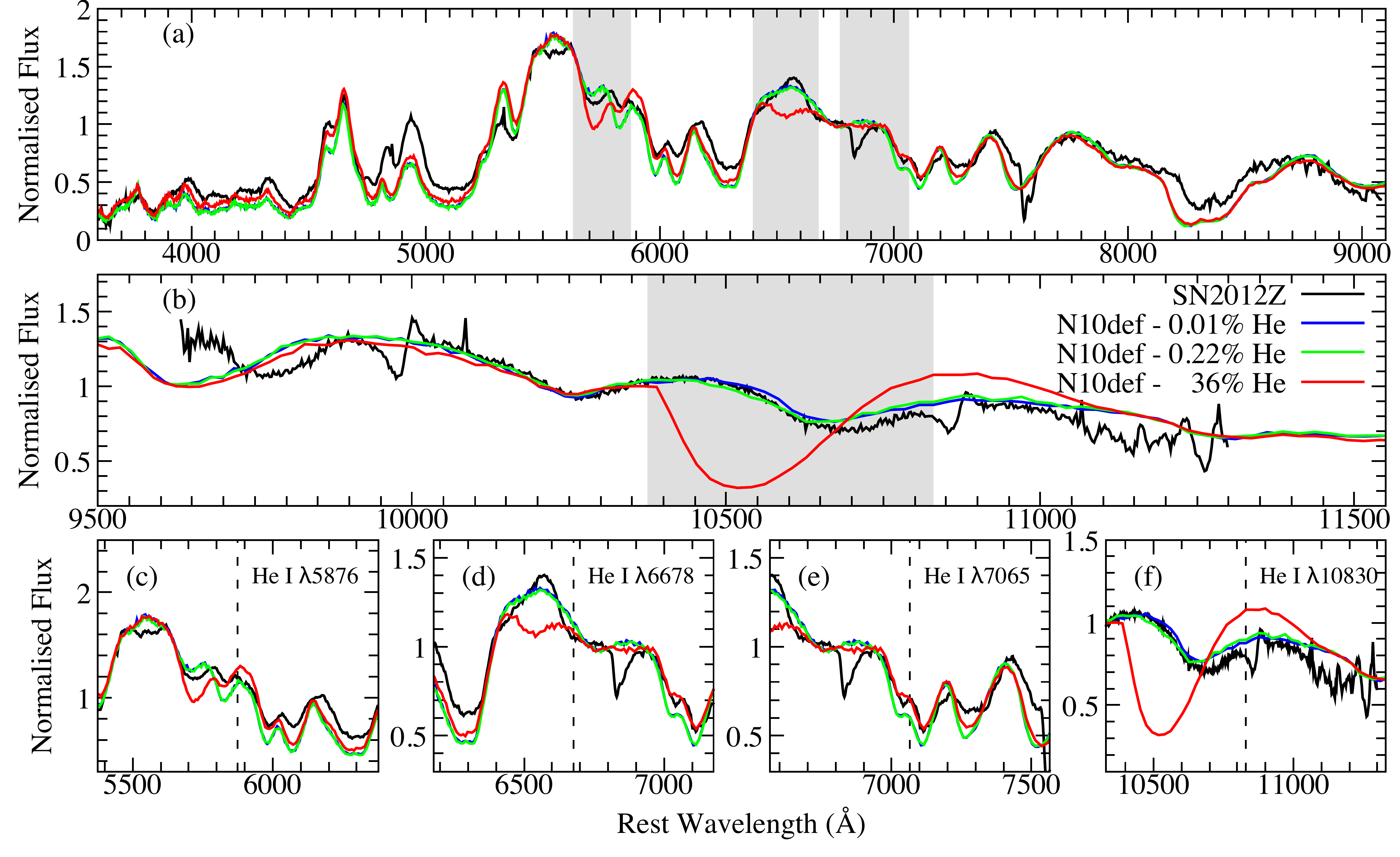}
\caption{Comparison of SN~20012Z and our N10def models with 0.01\%, 0.22\%, and 36\% helium abundances. Optical and infrared spectra have been normalised to the median flux from 5\,000\,\AA\, -- 7\,000\,\AA\, and 10\,000\,\AA\, -- 12\,000\,\AA, respectively. The optical and infrared spectra of SN~2012Z are approximately $+16$ days and $+17$ days after $B$-band maximum and have been corrected for galactic and host extinction. Our N10def model is shown at $+15$ days after bolometric maximum light. (a) and (b): Shaded regions represent \ion{He}{i}~$\lambda$5\,876, $\lambda$6\,678, $\lambda$7\,065, and $\lambda$10\,830 at rest and blue-shifted by the maximum velocity of the model ($\sim$12\,700~km~s$^{-1}$). (c), (d), (e) and (f): Zoom-ins of the regions surrounding \ion{He}{i}~$\lambda$5\,876, $\lambda$6\,678,  $\lambda$7\,065, and $\lambda$10\,830, respectively. Rest wavelengths are denoted by a vertical dashed line.}
\label{fig:12Z_helium_comparison}
\centering
\end{figure*}

In this section, we discuss comparisons between SN~2015H \cite[$M_{\rm{r}} = 17.27\pm0.07$~mag; ][]{15h}, SN~2005hk \cite[$M_{\rm{V}} = -18.08\pm0.29$~mag; ][]{phillips--07,kromer-13}, and SN~2012Z \cite[$M_{\rm{V}} = -18.50\pm0.09$~mag; ][]{comp--obs--12z} and the spectra generated from our N3def ($M_{\rm{V}} = -17.52$~mag), N5def ($M_{\rm{V}} = -18.24$~mag), and N10def ($M_{\rm{V}} = -18.38$~mag) models respectively. These are shown in Figs.~\ref{fig:15H_helium_comparison},~\ref{fig:05hk_helium_comparison}, and~\ref{fig:12Z_helium_comparison}. We consider these objects together as they show similar trends when compared to our synthetic spectra. Our N5def and N10def models show good agreement with SN~2005hk and SN~2012Z, reproducing many of the features observed with the appropriate velocity and strengths. This also holds for the \ion{Co}{ii} features at $\sim$16\,000\,\AA\, and from 22\,000\,\AA\, to 24\,000\,\AA ~in SN~2005hk. For SN~2015H, we find that the velocities in our N3def models are somewhat higher than those observed, however the model spectra successfully reproduce many of the features observed and the continuum shape.

\par

We find that increasing the helium abundance to 0.22\% has little effect on the optical spectra of any of these three models. Unlike the case of SN~2007J, our low helium abundance models (0.01\%, 0.22\%) show generally better agreement with the strengths of the \ion{Fe}{ii} absorption features centred around $\sim$5\,800~\AA\, and $\sim$5\,900~\AA~ in all three objects. As discussed in Sect.~\ref{sect:model_spectra}, the optical \ion{He}{i} features in more luminous models are generally weaker than in fainter models. Therefore, increasing the helium abundance further to 36\% is also generally consistent with these features in all three objects. Although the higher abundance helium models do produce optical \ion{He}{i} features, the level of agreement with the observations does not differ significantly. Based on the optical spectrum alone, it is difficult to exclude a large helium mass in all three cases.

\par

Increasing the helium abundance of these three models to 36\% produces additional \ion{He}{i} features that are inconsistent with observed spectra of each of the three supernovae. Each model produces a feature due to \ion{He}{i}~$\lambda$6\,678 that is not clearly observed in any object. The higher helium abundance models also produce strong \ion{He}{i}~$\lambda$10\,830 features that are clearly significantly stronger, and at higher velocities than the features observed in all three objects. Lower helium abundance models (0.01\%, 0.22\%) provide much better agreement and show that the features observed in SNe~2015H, 2005hk, and 2012Z could be the result of blended \ion{Fe}{ii}. In addition, the longer wavelength coverage of SN~2005hk allows us to show that the N5def model containing 36\% helium produces a strong absorption feature due to \ion{He}{i}~$\lambda$20\,587 that is also not observed in SN~2005hk.

\par

The N3def, N5def, and N10def models with 0.01\% or 0.22\% helium abundances do not produce strong features at any wavelength. We are therefore able to place an upper limit on the amount of helium present. SNe 2015H, 2005hk, and 2012Z are each consistent with no helium or mass fractions up to a few tenths of a percent, which is equivalent to \textless10$^{-3}~M_{\odot}$. Although a higher helium mass fraction could feasibly produce somewhat favourable agreement with certain optical features, the lack of a clearly identifiable \ion{He}{i}~$\lambda$10\,830 feature provides strong constraints on the helium content of these objects.

%

\subsection{SN~2010ae}

\begin{figure*}[!t]
\centering
\includegraphics[width=\textwidth]{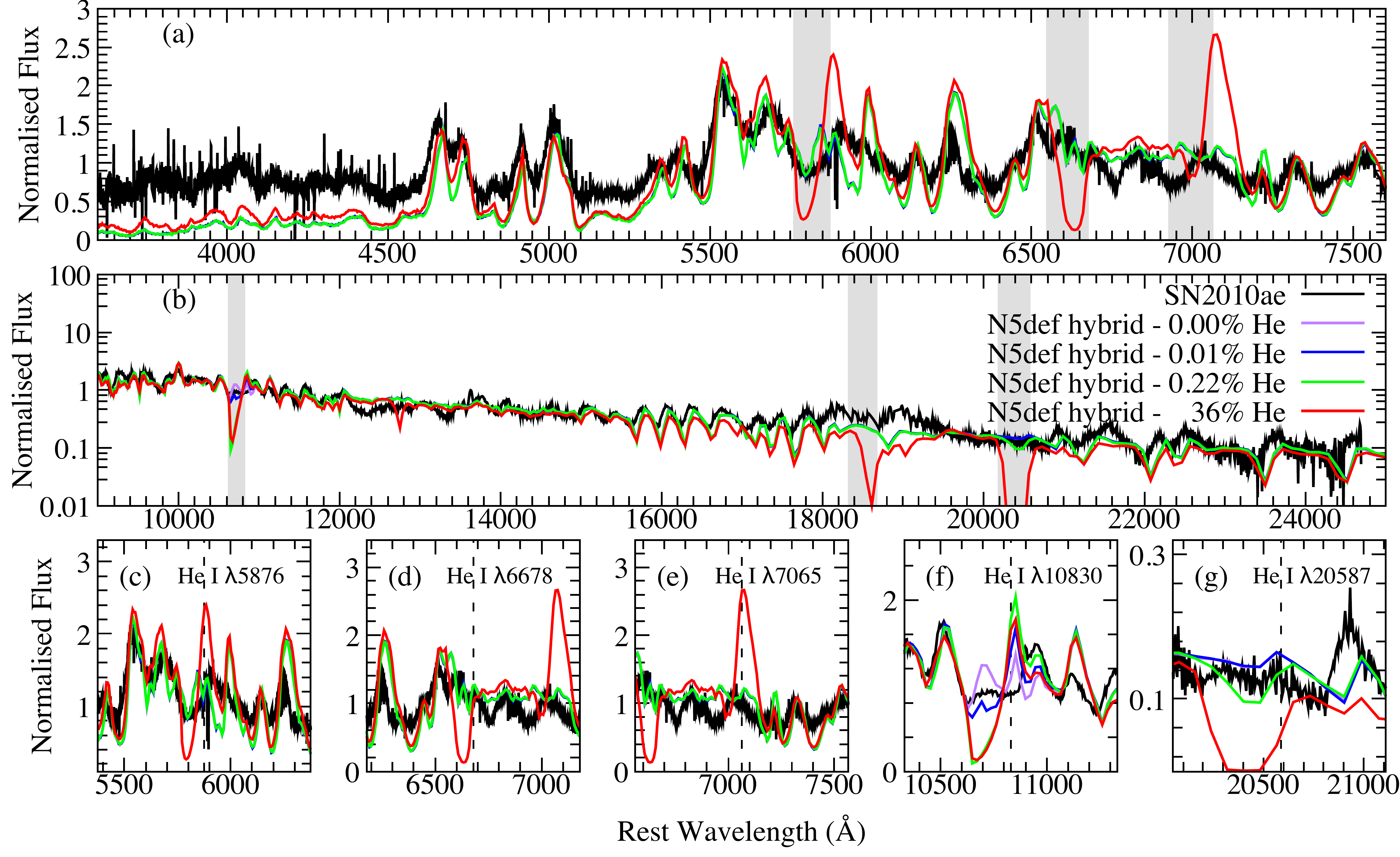}
\caption{Comparison of SN~2010ae and our N5def-hybrid models with 0\%, 0.01\%, 0.22\%, and 36\% helium abundances. Optical and infrared spectra have been normalised to the median flux from 5\,000\,\AA\, -- 7\,000\,\AA\, and 10\,000\,\AA\, -- 12\,000\,\AA, respectively. The optical and infrared spectra of SN~2010ae are approximately $+16$ days after $B$-band. The level of extinction experienced by SN~2010ae is highly uncertain \citep{obs--10ae}. Spectra presented here have been corrected for a total extinction of A$_{\rm{V}}$ = 1.9. Note that the infrared spectrum of SN~2010ae has been binned to $\Delta\lambda$ = 5~\AA. Our N5def-hybrid model is shown at $+15$ days after bolometric maximum light. (a) and (b): Shaded regions represent \ion{He}{i}~$\lambda$5\,876, $\lambda$6\,678, $\lambda$7\,065, $\lambda$10\,830, and $\lambda$20\,587 at rest and blue-shifted by the  maximum velocity of the model ($\sim$6\,000~km~s$^{-1}$). (c), (d), (e), (f), and (g): Zoom-ins of the regions surrounding \ion{He}{i}~$\lambda$5\,876, $\lambda$6\,678,  $\lambda$7\,065, $\lambda$10\,830, and $\lambda$20\,587, respectively. Rest wavelengths are denoted by a vertical dashed line. 
}
\label{fig:10ae_helium_comparison}
\centering
\end{figure*}

SN~2010ae is one of the least luminous SNe Iax to be discovered to date \cite[$-15.33\lesssim M_{\rm{V}} \lesssim -13.80\pm0.54$~mag][]{obs--10ae} and is spectroscopically similar to SN~2008ha. The N5def-hybrid model ($M_{\rm{V}} = -14.52$~mag) has been previously shown to produce spectra that are broadly consistent with the observed properties of SN~2008ha \citep{kromer-15}. In Fig.~\ref{fig:10ae_helium_comparison}, we show a comparison between our N5def-hybrid models and SN~2010ae. We find that models with an inner boundary velocity of 1100~km~s$^{-1}$ produce synthetic spectra that best match SN~2010ae. These models are able to reproduce many, but not all, of the narrow features observed in the spectra of SN~2010ae. We note that while the spectrum of SN~2010ae presented in this section has been corrected for $E(B-V)_{\rm{tot}} = 0.62$~mag, the extinction of SN~2010ae is highly uncertain \cite[$E(B-V)_{\rm{tot}} \sim 0.62\pm0.42$~mag;][]{obs--10ae}. 

\par

Figure~\ref{fig:10ae_helium_comparison} shows that the spectra of SN~2010ae are inconsistent with a helium mass constituting 36\% of the ejecta. The N5def-hybrid model with a helium abundance of 36\% produces optical \ion{He}{i} features that are too strong to match SN~2010ae. In addition to the strong \ion{He}{i}~$\lambda$10\,830 and $\lambda$20\,587 features, this model shows strong \ion{He}{i}~$\lambda$18\,685, a feature that is clearly inconsistent with SN~2010ae at these wavelengths. 

\par

In Fig.~\ref{fig:10ae_helium_comparison} we include a synthetic spectrum from the N5def-hybrid model with 0\% helium. For more luminous models, we are generally unable to distinguish between models with 0.22\% and 0.01\% helium, and hence also no helium. Figure~\ref{fig:10ae_helium_comparison} clearly shows however, that the low helium abundance models produce distinct infrared features. Figure~\ref{fig:10ae_helium_comparison}(f) shows that the models containing 36\% and 0.22\% helium produce \ion{He}{i}~$\lambda$10\,830 that is much too strong to match SN~2010ae. Our model containing no helium predicts a double peaked feature in this region due to blended \ion{Fe}{ii} lines. In contrast, the helium contribution from our 0.01\% helium model produces a broader, flat-bottomed feature that is more similar to SN~2010ae. At all other wavelengths our models containing no and 0.01\% helium are indistinguishable. Therefore, we speculate that SN~2010ae could contain a helium abundance on the order of a few hundredths of a percent. 

\par

By fitting the pseudo-bolometric light curve of SN~2010ae with Arnett's law \citep{arnett--law}, \cite{obs--10ae} estimate the ejecta mass of SN~2010ae to be $\sim$0.3 -- 0.6~$M_{\odot}$. The N5def-hybrid ejecta mass is substantially lower \cite[0.014~$M_{\odot}$;][]{kromer-15}. If an explosion model could produce spectra similar to the N5def-hybrid model, but also contain a larger ejecta mass, then it would be possible that SN~2010ae could contain a few $10^{-6}$ -- 10$^{-4}~M_{\odot}$ of helium. We note that the \ion{Fe}{ii} features centred around $\sim$10\,700\,\AA\, are likely blended in our higher luminosity models. It is possible that if these features were blended in our N5def-hybrid model, they would also provide a good match with the feature observed in SN~2010ae and we would be unable to discern the helium contribution. Our model shows good agreement with many of the narrow features observed, such as the \ion{Co}{ii} features at $\sim16\,000$\,\AA, indicating that more blending would decrease the level of overall agreement.

%

\subsection{SN~2014ck}

\begin{figure*}[!t]
\centering
\includegraphics[width=\textwidth]{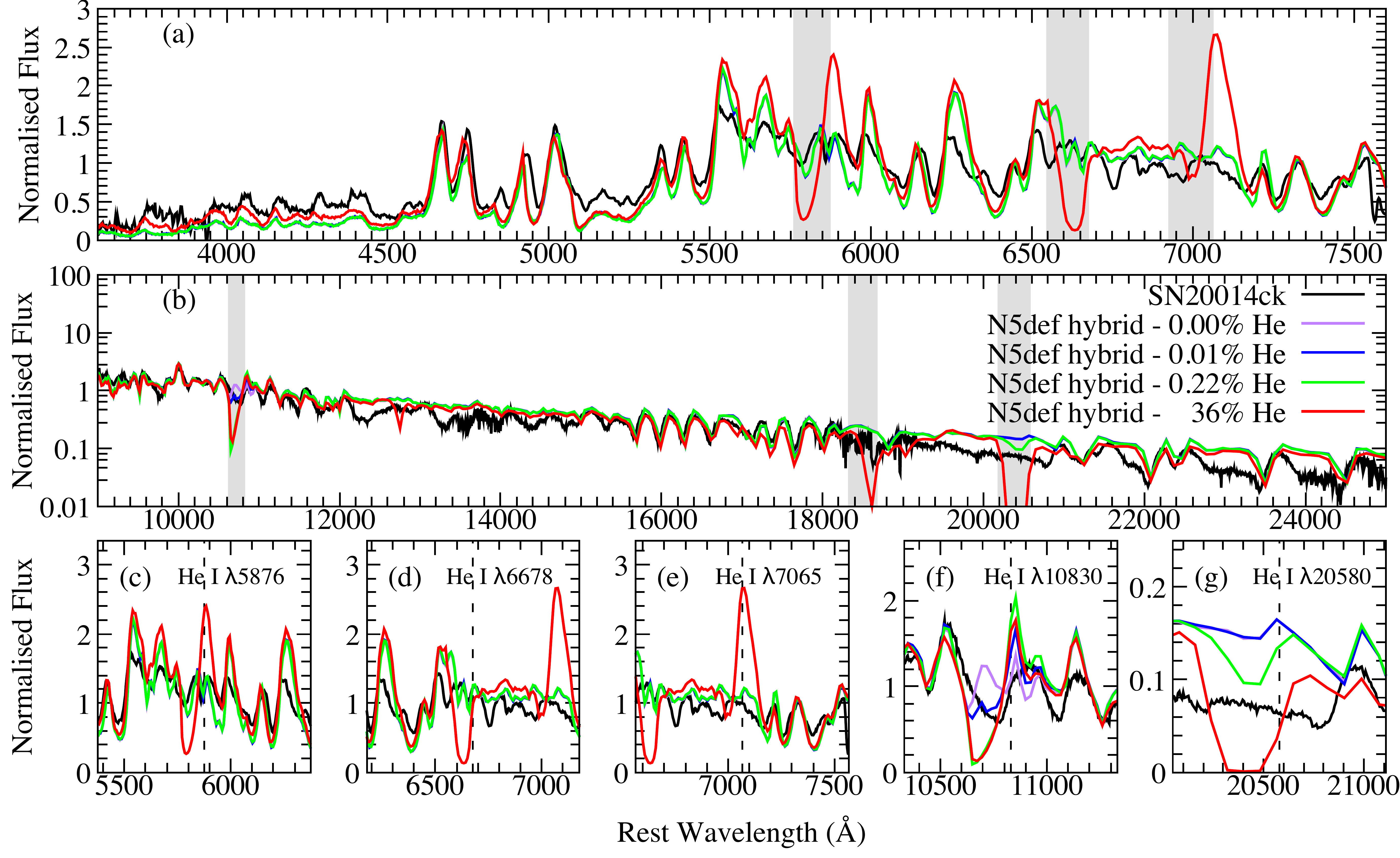}
\caption{Comparison of SN~2014ck and our N5def-hybrid models with 0\%, 0.01\%, 0.22\%, and 36\% helium abundances. Optical and infrared spectra have been normalised to the median flux from 5\,000\,\AA\, -- 7\,000\,\AA\, and 10\,000\,\AA\, -- 12\,000\,\AA, respectively. The optical and infrared spectra of SN~2014ck are approximately $+17$ and $+19$ days after $V$-band. Our N5def-hybrid model is shown at $+15$ days after bolometric maximum light. (a) and (b): Shaded regions represent \ion{He}{i}~$\lambda$5\,876, $\lambda$6\,678, $\lambda$7\,065, $\lambda$10\,830, and $\lambda$20\,587 at rest and blue-shifted by the  maximum velocity of the model ($\sim$6\,000~km~s$^{-1}$). (c), (d), (e), (f), and (g): Zoom-ins of the regions surrounding \ion{He}{i}~$\lambda$5\,876, $\lambda$6\,678,  $\lambda$7\,065, $\lambda$10\,830, and $\lambda$20\,587, respectively. Rest wavelengths are denoted by a vertical dashed line.}
\label{fig:14ck_helium_comparison}
\centering
\end{figure*}

Although the light curve of SN~2014ck is similar to SN~2015H in terms of peak magnitude and decline rate ($M_{\rm{V}} = -17.29\pm0.15$~mag), \cite{tomasella--2016} show that
the spectra are dominated by low velocity features that are most similar to SN~2008ha. This presents a challenge for the model sets explored in this work \cite[i.e. the explosion models of][]{fink-2014}, as they show a general correlation between peak luminosity and velocity. We note however, that the \cite{fink-2014} models were not a complete exploration of the parameter space for pure deflagrations. It currently remains to be seen whether it is possible to produce a model with a luminosity similar to N3def \cite[$M_{\rm{V}}$ = $-$17.52~mag;][]{fink-2014}, but spectral features similar to the N5def-hybrid model ($M_{\rm{V}} = -14.52$~mag). Nevertheless, we have compared the spectra of SN~2014ck to synthetic spectra generated from our N1def, N3def, and N5def-hybrid models. We find that our N1def and N3def models produce features much broader and at higher velocities than those observed in SN~2014ck, while the N5def-hybrid shows better agreement (see Fig.~\ref{fig:14ck_helium_comparison}). We caution that while the N5def-hybrid model does show good agreement spectroscopically, this model is approximately three magnitudes fainter at peak and shows a much faster decline rate than SN~2014ck. We therefore explore the spectral features but do not attempt to present helium mass estimates. 

\par

In Fig.~\ref{fig:14ck_helium_comparison} we show that the narrow spectral features of the N5def-hybrid model provide generally good agreement with SN~2014ck. Again, the faintness of this model allows for strong \ion{He}{i} features to form and Fig.~\ref{fig:14ck_helium_comparison}(a) and (b) show that our model with 36\% helium produces strong P-Cygni profiles that are in clear disagreement with the observed spectra. Unlike SN~2010ae, the broad feature around $\sim$10\,800~\AA\, does not show a flat bottom. Clearly this feature is also in disagreement with the two narrower \ion{Fe}{ii} features predicted by our model containing no helium. 

\par

Future studies should explore scenarios in which models can produce relatively bright explosions, but also show narrow velocities. A model producing a substantially larger ejecta mass, while maintaining the same kinetic energy and $^{56}$Ni mass, could provide an avenue to explain the peculiarities of SN~2014ck. 

%

\section{Discussion}
\label{sect:discussion}

\subsection{Limitations of the models}
\label{sect:model_limits}

As discussed by \cite{boyle--17}, the most important limiting factor for the approximations used as part of this study is whether helium remains sufficiently ionised during the epochs investigated. If the true degree of ionisation is lower than assumed in our modelling, then the strengths of the helium features presented in this work may be taken to be upper limits, which would result in an underestimation of the helium masses \citep{boyle--17}. The pure deflagration models of \cite{fink-2014} that are used in this work have been shown to produce an ejecta mass lower than those predicted for SNe Iax \citep{kromer-13, kromer-15, 15h} and could therefore also lead to an underestimation of the helium mass. This ranges from a factor of approximately a few, to as much $\gtrsim$20 in the case of SN~2010ae and N5def-hybrid. 

\par

The relative importance of these uncertainties, or indeed whether one is dominant over the other, is unclear. We therefore strongly advise against using the helium mass fractions presented in this work and the ejecta masses of the \cite{fink-2014} pure deflagration models to obtain an absolute helium mass. Instead, we provide qualitative mass estimates, the uncertainties on which should be assumed to be a factor of a few. We note however, that while there may be some uncertainty on the absolute masses, the general trends should still hold true.

\subsection{Sources of helium in the context of carbon deflagrations}
\label{sect:helium_source}

Our models suggest that all SNe Iax within our sample are consistent with at least some amount of helium in their ejecta, although we stress that only one object (SN~2007J) definitively requires the presence of helium. If this helium originated from a helium star companion, it could have been unbound from the companion due to the supernova explosion itself. The impact of supernova ejecta on a companion star in a tight binary can lead to a significant amount of mass being removed from the companion, depending on the nature of the companion and orbital properties \citep{marietta--2000, pakmor--2008, pan--2010}.

\par

\cite{pan--2012} find that the amount of hydrogen that SN Ia ejecta could strip from a main sequence companion is typically $\lesssim$0.2~$M_{\odot}$. \cite{liu--2013a} show that the amount of material stripped correlates strongly with explosion energy and \cite{liu--2013b} show specifically that the less energetic explosions of SNe Iax could strip only $\sim$0.01 -- 0.02~$M_{\odot}$ of hydrogen from a main sequence companion - $\sim$10\% of the mass stripped by SNe Ia. Helium main sequence star donors are more compact and therefore the amount of helium stripped is approximately an order of magnitude smaller than the amount of hydrogen stripped from a main sequence companion \citep{pan--2012}. This suggests that SNe Iax could only strip $\lesssim$0.002~$M_{\odot}$ from a helium main sequence star. If the companion was a more evolved and extended star, such as a helium sub-giant, more material could be stripped. 

\par

Our modelling implies that the helium mass in SN~2007J could be as high as a few 10$^{-2}~M_{\odot}$, which appears too large to have been stripped from a helium main sequence companion, but could be comparable to the amount stripped from an evolved helium star companion. An important prediction for stripped material however, is that it should be concentrated at low velocities \cite[$\lesssim$1\,000~km~s$^{-1}$;][]{marietta--2000}. The spectra of SN~2007J show no indication that helium exists at a significantly different velocity from the rest of the ejecta, therefore the helium present in the ejecta is unlikely to have been stripped from a companion. For SN~2010ae, we find good agreement with the N5def-hybrid model. The kinetic energy of this model is approximately 100 times lower than the model used by \cite{liu--2013b} and therefore could plausibly strip only $\sim10^{-5}~M_{\odot}$ of helium from a companion. As noted previously, the ejecta mass for the N5def-hybrid model is significantly lower than the estimated ejecta mass of SN~2010ae - therefore the kinetic energy of SN~2010ae is likely substantially higher than the N5def-hybrid model. Nevertheless, the tentative helium mass estimate for SN~2010ae (a few 10$^{-6}$ -- 10$^{-4}~M_{\odot}$) is broadly consistent with stripping from a helium star companion. The upper limits derived for SNe 2005hk, SN~2012Z, and SN~2015H ($\textless10^{-3}~M_{\odot}$), which are based on non-detections of helium features, are also consistent with stripping. 

\par

Alternatively, accretion of helium-rich material from a companion star could be a viable scenario to explain helium present in SN Iax ejecta. \citep{bildsten--07,neunteufel--2016,wang--2017}. This would require there to be unburned helium present at the onset of the explosion in the core. For low accretion rates ($\lesssim$10$^{-6}$~\.{M}$_{\odot}$~yr$^{-1}$) the accreted helium forms a shell and may eventually ignite unstable burning - in which the shell undergoes periodic helium flashes \citep{shen--09}. At higher accretion rates, helium is burned directly into carbon \citep{iben--89}. In the pure deflagration scenario, the deflagration front does not burn the entirety of the star. This could result in at least some amount of the helium shell remaining intact after the explosion and becoming mixed within the freely expanding ejecta.

\par

The mass required to plausibly ignite the helium shell decreases with increasing white dwarf mass \citep{shen--09}. The helium mass estimated for SN~2007J (a few 10$^{-2}~M_{\odot}$) is likely too large for a helium shell accreted onto a Chandrasekhar-mass white dwarf. Perhaps a sufficiently large helium shell mass ($\sim 0.1$ -- $0.3~M_{\odot}$) could be achieved in systems with low mass helium donors ($\sim$0.5~$M_{\odot}$) in which the accretion rate is driven by gravitational wave radiation \cite[see e.g][]{nelemans--01}. The delay time for such a system however, would likely be much longer than that expected for SNe Iax ($\lesssim$100~Myr). The low helium masses estimated for all other objects in our sample are consistent with accretion from a helium star donor.

\subsection{Alternative scenarios}
\label{sect:alternatives}
Our modelling shows that SN~2007J requires a relatively large helium mass to reproduce the features observed while all other objects in our sample are consistent with containing either no helium or a small helium mass. We are therefore unable to determine whether the amount of helium present in SNe Iax follows a continuous distribution or if there is a bi-modal distribution in which some objects contain large helium abundances while others contain no helium. 

\par

SN~2007J represents an outlier among SNe Iax and therefore may not share a common origin. As discussed previously, the large helium mass required by SN~2007J appears difficult to reconcile with current expectations for SNe Iax and helium star companions. Alternative scenarios include the possibility that SN~2007J was not the pure deflagration of a carbon-oxygen core in a Chandrasekhar-mass white dwarf. If instead SN~2007J resulted from a pure helium shell deflagration, this would produce a small amount of radioactive $^{56}$Ni and a significant amount of unburned helium. \cite{woosley--11} present a series of one-dimensional explosion models for various white dwarf masses and accretion rates. Those models which undergo helium deflagrations eject similar amounts of mass as carbon deflagration explosions ($\sim$0.08 -- 0.14~$M_{\odot}$). Most of this ejecta was unburned helium with only a small amount of $^{56}$Ni ($\sim$6 -- 9 $\times 10^{-5}$~$M_{\odot}$) produced, while substantial amounts of $^{44}$Ti and $^{48}$Cr were synthesised. Full multi-dimensional simulations are necessary to fully capture the mixing produced by deflagrations and test whether helium deflagrations can provide a viable scenario for SN~2007J.

\par

Finally, we also note the possibility that SN~2007J is not a genuine SN Iax. As mentioned in Sect.~\ref{sect:intro}, \cite{slow--ptf} argue that SN~2007J is a SN IIb, similar to SN~1993J. \cite{foley--late--iax} contradict this claim and instead argue that the H$\alpha$ feature identified by \cite{slow--ptf} is similar to a feature in SN~2002cx identified as Fe II by \cite{read--02cx--spectra}. Nevertheless, SN~2007J does share spectral similarities to SN~I/IIb and while \cite{foley--late--iax} argue that it is most similar to SNe Iax, they do concede that it may not be physically related. If SN~2007J cannot be reconciled with a thermonuclear origin, alternative scenarios should be explored.

%

\section{Conclusions}
\label{sect:conclusions}

In this study, we presented a series of spectra for pure deflagration models of carbon-oxygen and hybrid carbon-oxygen-neon white dwarfs containing varying amounts of helium. Our models show that the helium spectral features are stronger in less luminous models and increase in strength over time for all models. In addition, \ion{He}{i}~$\lambda$10\,830\, is typically the strongest \ion{He}{i} feature produced - the optical features of \ion{He}{i} are only produced by models with large helium abundances ($\gtrsim$10\%) and are always accompanied by strong \ion{He}{i}~$\lambda$10\,830\, absorption. Our results demonstrate that infrared and post-maximum spectra, and fainter objects, provide the best opportunities to test for the presence of helium in SNe Iax.

\par

We compared our model spectra to SN~2007J, which is argued to be a SN Iax showing \ion{He}{i} spectral features. SN~2007J did not have an infrared spectrum and therefore we are unable to test the strength of \ion{He}{i}~$\lambda$10\,830. We find that the optical spectral features of SN~2007J are consistent with our N1def model (peak $M_{\rm{V}} = -16.8$~mag) containing a large helium abundance (on the order of tens of percent, equivalent to a few $10^{-2}~M_{\odot}$). We argue that current models for accretion and material stripping from a helium star companion struggle to produce a compatible scenario to SN~2007J. If the large helium mass invoked for SN~2007J can not be reconciled with thermonuclear scenarios, alternatives should be investigated further.

\par

We also compared our models to all existing SNe Iax with infrared spectra. We find that SNe~2005hk, 2012Z, and 2015H show no signs of strong \ion{He}{i} features and are therefore consistent with helium abundances of  $\textless$10$^{-3}~M_{\odot}$. For SN~2010ae, we find good agreement with our N5def-hybrid model (peak $M_{\rm{V}} = -14.52$~mag) containing a small amount of helium. Due to the differences in ejecta mass between our model and SN~2010ae, this is likely equivalent to a range of 10$^{-6}$ -- $10^{-4}~M_{\odot}$ of helium. The helium abundances for these objects are consistent with either stripping or accretion from a helium star companion.

\par

Currently there is no evidence to suggest that a correlation exists between the amount of helium present and other supernova properties, such as peak brightness. We are limited however, by the small sample of objects with available infrared observations. Obtaining infrared spectra of SNe Iax should become a focus of follow-up campaigns on future objects, as the presence and amount of helium in the ejecta provides important constraints for the progenitor system. 

%

\begin{acknowledgements}
We thank the anonymous referee for their careful consideration of our manuscript. We thank M. Kromer and L. Tomasella for providing spectra of SNe 2005hk and 2014ck, respectively, and A. Ruiter and K. Shen for useful and informative discussions. MM acknowledges funding and support from the University of Turku. MM and KM are funded by the EU H2020 ERC grant no.~758638. SS and KM acknowledge support from STFC through grant ST/P000312/1. This work made use of the Queen's University Belfast computing cluster, Kelvin, and the Heidelberg Supernova Model Archive (HESMA), https://hesma.h-its.org. This research has made use of the Keck Observatory Archive (KOA), which is operated by the W. M. Keck Observatory and the NASA Exoplanet Science Institute (NExScI), under contract with the National Aeronautics and Space Administration.
\end{acknowledgements}

\bibliographystyle{aa}
\bibliography{Magee_SNeIax+He}

\begin{appendix}

\onecolumn
\section{Model spectra}
\begin{figure}[!h]
\centering
\includegraphics[width=15cm]{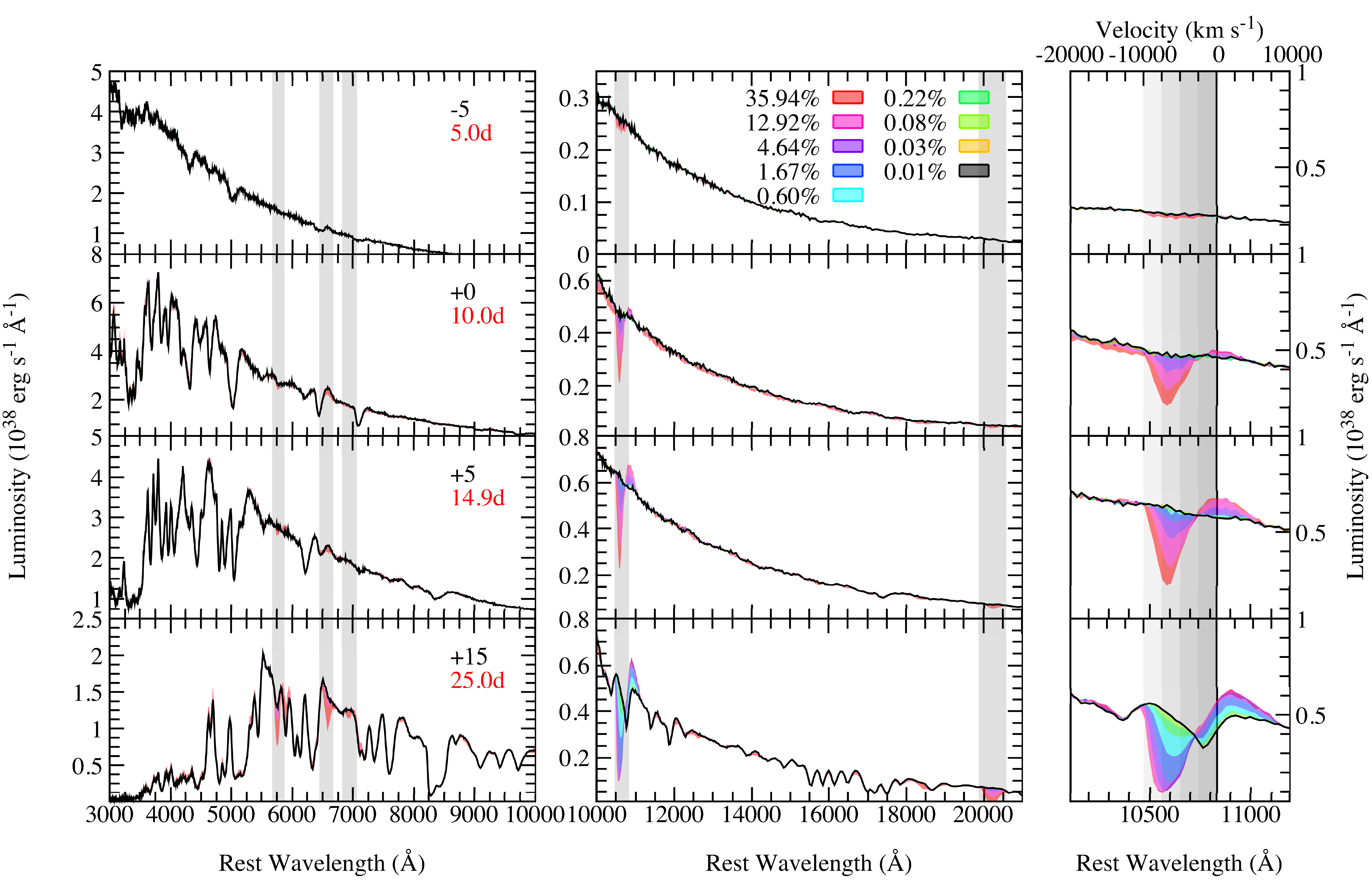}
\caption{N3def model spectra at various epochs, with varying helium abundances given as percentages of ejecta mass. Shaded regions show the difference produced in the spectra by varying the helium abundance. Phases relative to bolometric maximum light are given in black, while times since explosion are given in red. {\it Left:} Grey shaded regions represent \ion{He}{i}~$\lambda$5\,876, $\lambda$6\,678, $\lambda$7\,065, $\lambda$10\,830, and $\lambda$20\,587 at rest and blue-shifted by the maximum velocity of the model ($\sim$10\,400~km~s$^{-1}$). {\it Right:} Zoomed in regions surrounding \ion{He}{i}~$\lambda$10\,830. Each shaded region corresponds to a blue-shift of 2\,500~km~s$^{-1}$.
}
\label{fig:n3def_varying_helium}
\centering
\end{figure}

\begin{figure}[!h]
\centering
\includegraphics[width=15cm]{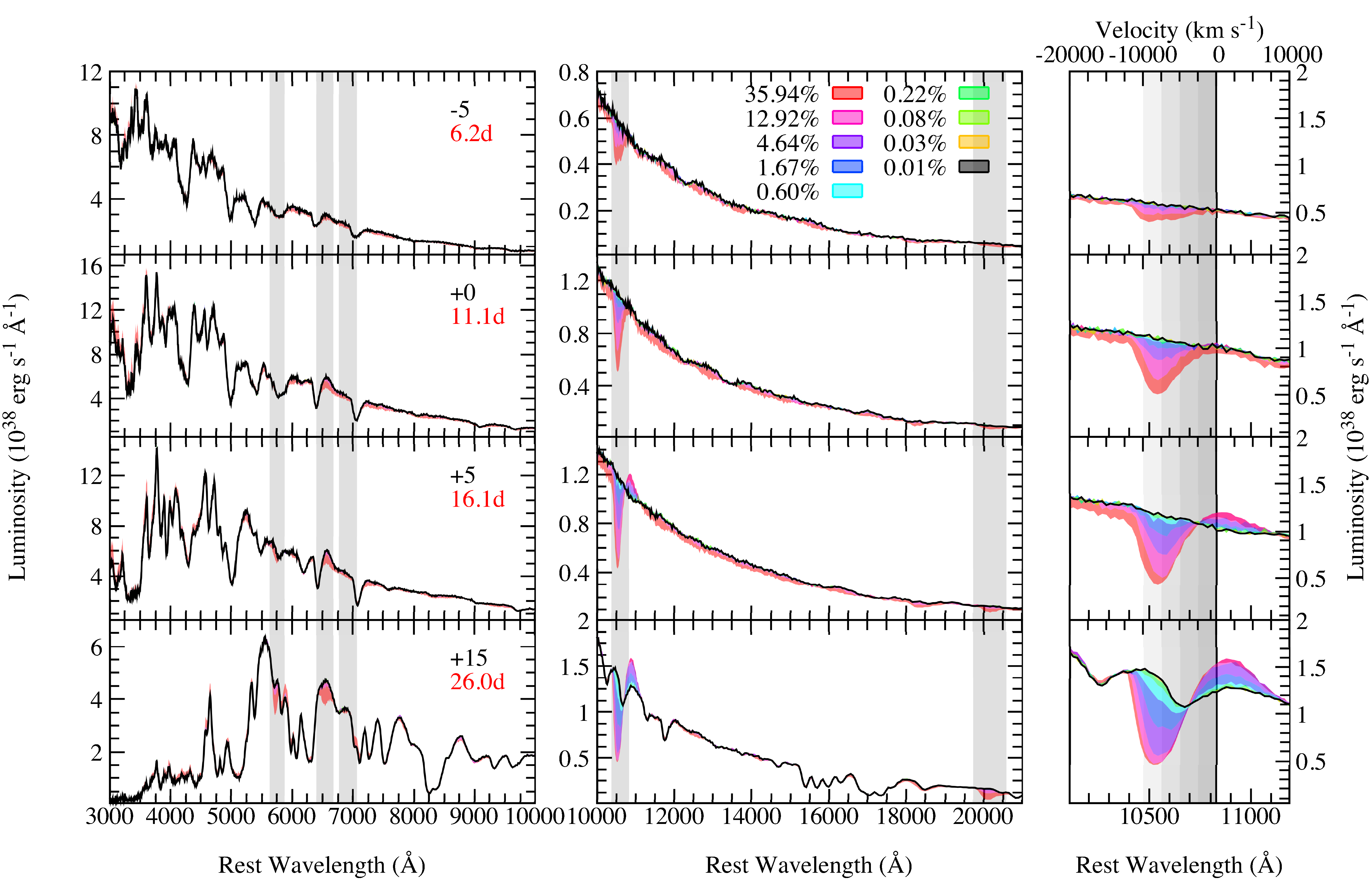}
\caption{N10def model spectra at various epochs, with varying helium abundances given as percentages of ejecta mass. Shaded regions show the difference produced in the spectra by varying the helium abundance. Phases relative to bolometric maximum light are given in black, while times since explosion are given in red. {\it Left:} Grey shaded regions represent \ion{He}{i}~$\lambda$5\,876, $\lambda$6\,678, $\lambda$7\,065, $\lambda$10\,830, and $\lambda$20\,587 at rest and blue-shifted by the maximum velocity of the model ($\sim$12\,700~km~s$^{-1}$). {\it Right:} Zoomed in regions surrounding \ion{He}{i}~$\lambda$10\,830. Each shaded region corresponds to a blue-shift of 2\,500~km~s$^{-1}$.
}
\label{fig:n10def_varying_helium}
\centering
\end{figure}

\begin{figure}
\centering
\includegraphics[width=15cm]{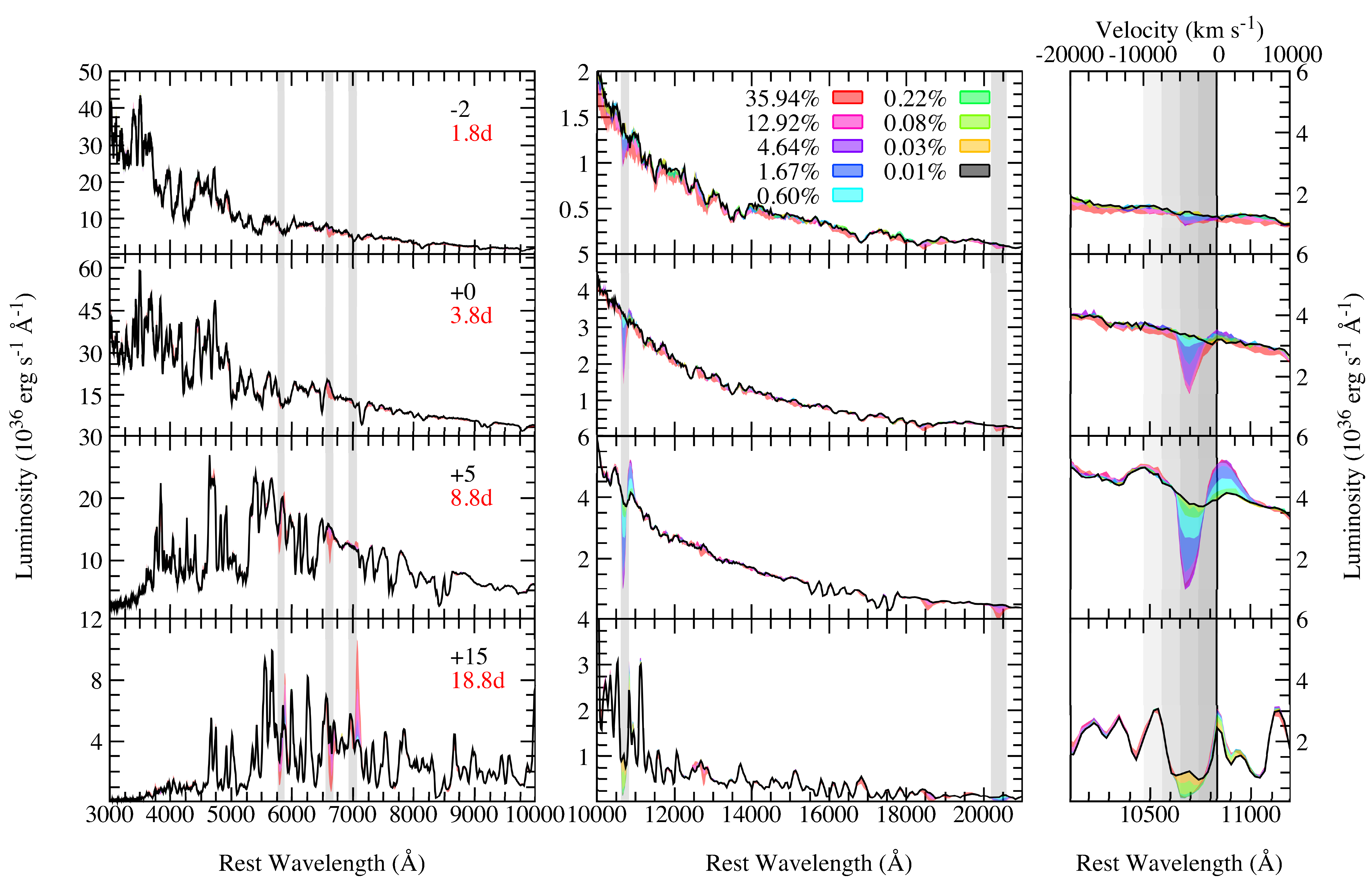}
\caption{N5def-hybrid model spectra at various epochs, with varying helium abundances given as percentages of ejecta mass. Shaded regions show the difference produced in the spectra by varying the helium abundance. Phases relative to bolometric maximum light are given in black, while times since explosion are given in red. {\it Left:} Grey shaded regions represent \ion{He}{i}~$\lambda$5\,876, $\lambda$6\,678, $\lambda$7\,065, $\lambda$10\,830, and $\lambda$20\,587 at rest and blue-shifted by the maximum velocity of the model ($\sim$6\,000~km~s$^{-1}$). {\it Right:} Zoomed in regions surrounding \ion{He}{i}~$\lambda$10\,830. Each shaded region corresponds to a blue-shift of 2\,500~km~s$^{-1}$.
}
\label{fig:n5def_hybrid_varying_helium}
\centering
\end{figure}

\end{appendix}

\end{document}